\documentclass{article}[12pt,a4paper]
\usepackage{graphicx}  
\usepackage{dcolumn}   
\usepackage{bm}        
\usepackage{amssymb}   
\usepackage[]{amsmath,amssymb}
\usepackage{graphics,epsfig}
\usepackage[latin1]{inputenc}
\usepackage[T1]{fontenc}
\usepackage{amsfonts}
\usepackage{latexsym}
\usepackage{physics}
\usepackage[english]{babel}
\usepackage{subcaption} 
\newcommand{\be}{\begin{equation}}
\newcommand{\ee}{\end{equation}}
\newcommand{\beq}{\begin{equation}}
\newcommand{\eeq}{\end{equation}}
\newcommand{\bea}{\begin{eqnarray}}
\newcommand{\eea}{\end{eqnarray}}
\newcommand{\nn}{\nonumber}
\def\be{\begin{equation}}
\def\ee{\end{equation}}
\def\ba{\begin{eqnarray}}
\def\ea{\end{eqnarray}}

\numberwithin{equation}{section}

\begin{document}


\title{Anisotropic Unruh temperatures}

\author{Ra\'{u}l E. Arias$^{a}$, Horacio Casini$^b$, Marina Huerta$^b$, Diego Pontello$^b$}

\maketitle

\begin{center}

{\sl $^a$ Instituto de F\'{\i}sica de La Plata - CONICET\\
C.C. 67, 1900 La Plata, Argentina.}

~

{\sl $^b$ Centro At\'omico Bariloche,\\
8400-S.C. de Bariloche, R\'{\i}o Negro, Argentina}

\end{center}

\begin{abstract}
The relative entropy between very high energy localized excitations and the vacuum, where both states are reduced to a spatial region, gives place to a precise definition of a local temperature produced by vacuum entanglement across the boundary. This generalizes the Unruh temperature of the Rindler wedge to arbitrary regions. The local temperatures can be read off from the short distance leading terms in the modular Hamiltonian. For free scalar and fermion fields they have a universal geometric expression that follows by solving a particular eikonal type equation in Euclidean space. This equation generalizes to any dimension the holomorphic property that holds in two dimensions. For regions of arbitrary shapes the local temperatures at a point are direction dependent. We compute their explicit expression for the geometry of a wall or strip.      
\end{abstract}

\section{Introduction}
Recently there has been a growth of interest in the study of modular Hamiltonians (see for example \cite{cardy,otros1,otros2,otros3,otros4,otros5}). 
The importance of this operator is due to its use in different disciplines of theoretical physics like quantum information, quantum field theory, condensed matter and quantum gravity. One of the applications to quantum information arises in the fact the modular Hamiltonian is related to the relative entropy, which measures the distinguishability between quantum states. Moreover, in condensed matter, for example, its spectrum is useful to identify topological states \cite{topo}, and from the quantum gravity point of view the modular Hamiltonian is involved in the first law of entanglement \cite{first}, which is a generalization of the first law of thermodynamics to non-equilibrium systems. It was shown that from the first law one can obtain the linear Einstein equations in a holographic theory \cite{first-einstein,first-einstein2}. In addition, the modular Hamiltonian was used to understand the Bekenstein bound as well as prove other entropy bounds arising from black hole physics \cite{bounds,bounds2,bounds3,bounds4,bounds5,bounds6,bounds7,bounds8}. In quantum field theory, it is part of the Tomita-Takesaki theory that plays an important role in the structure of the algebraic axiomatic approach \cite{haag,revi}. 

The modular Hamiltonian ${\cal H}_V$ arises from writing the reduced density matrix $\rho_V$ of the vacuum on a region $V$, in a thermal-like form, 
\be
\rho_V=c\, e^{-{\cal H}_V}\,.
\ee
 Remarkably, in \cite{Bisognano} it was shown that for any relativistic quantum field theory, and when the region $V$ is half spatial plane, the modular Hamiltonian is proportional to the stress energy tensor
\be
{\cal H}_V=2\pi\int_V d^{d-1}x\,  x_1\, T_{00}(x).\label{RindlerWedge}
\ee
Due to the Killing symmetry of the Rindler wedge (the domain of dependence of half-space) the expression is completely local and the unitaries $e^{i\,\tau\,{\cal H}_V}$ generate a local modular flow of field operators, which in this case coincides with boost transformations.   
This same expression (\ref{RindlerWedge}) is the reason behind the Unruh temperature for accelerated observers \cite{unruh}. These observers have time evolution dictated by the boost generator and accordingly can identify the vacuum state as a thermal state with respect to this evolution operator.   

In \cite{nosotros} a generalization of Unruh temperature suitable to arbitrary regions was proposed. This comes from a reinterpretation of Unruh temperature in terms of the relative entropy, and is directly related to the structure of local terms in the modular Hamiltonian. The relative entropy between two states  $\rho_V$, $\rho_V^1$ (in the same region) is given by 
\be
S_{\textrm{rel}}(\rho_V^1||\rho_V)=\Delta \langle {\cal H}_V\rangle- \Delta S\,,   
\ee
where $\Delta \langle {\cal H}\rangle=\textrm{tr}(\rho_V^1{\cal H}_V )-\textrm{tr}(\rho_V {\cal H}_V)$ is the variation in expectation values of the modular Hamiltonian and $\Delta S=S^1-S$ is the difference in the entanglement entropies. Taking as $\rho^1$ a localized high-energy excitation above the vacuum state around a point $x\in V$ the difference in entanglement entropies is negligible with respect to $\Delta \langle {\cal H}\rangle$, and the relative entropy in half-space is given by
\be
S_{\textrm{rel}}\sim \Delta \langle {\cal H}_V\rangle\sim \beta(x)  E\,,   
\ee
where $E$ is the excitation energy and 
\be
\beta(x)=2 \pi x_1
\ee
is the coefficient of $T_{00}$ in (\ref{RindlerWedge}); it coincides with the inverse Unruh temperature for the accelerated observer passing through the point with coordinate $x_1$. 

Relative entropy measures the distinguishability between states. More precisely, the probability of confounding the two states after $N$ judiciously chosen experimental measures declines with $N$ as $e^{- S_{\textrm{rel}} N}$ \cite{relative}. In the present case, one can roughly say that there is a probability $\sim e^{-S_{\textrm{rel}}}$
 of finding the excitation in the vacuum. As the entropy is small, this is given by the Boltzmann factor $e^{-{\cal H}}$ corresponding to the modular Hamiltonian.  
 The measurements to differentiate the two states have to be done inside the region $V$, including its causal domain of dependence. In this sense, Unruh temperature arises because the vacuum cannot be perfectly distinguished from the excitation with a finite number of measurements involving the restricted class of operators at disposal in $V$.\footnote{See \cite{carnot} for another interesting interpretation of Unruh temperature in terms of Carnot efficiencies.} This is a direct consequence of entanglement between $V$ and the rest of the space.  
The Unruh-de Witt accelerated detector interacting with vacuum fluctuations is a particular convenient and optimal experiment that one can do inside the Rindler wedge to measure this temperature. For other regions one has to resort to other kind of experiments to measure relative entropy.     

In \cite{nosotros} it was noticed that relative entropy monotonicity implies that for a general region and quantum field theory we also have an analogous expression for the relative entropy between the vacuum and a localized high momentum excitation
\be
 S_{\textrm{rel}}\sim \Delta \langle {\cal H}_V\rangle\sim \beta_V(x, \hat{p})  E\,.
\ee
That is, the relative entropy scales with the energy of the excitation, but now, in contrast to the case of the Rindler wedge, we have an inverse temperature $\beta$ that might also depend on the direction of the excitation momentum. In this case, the local structure of the modular Hamiltonian cannot be given exclusively in terms of the energy density operator, while still has to have the same scaling dimensions.\footnote{For free fields, the local term with the right scaling dimension is obtained by integrating two fields at two points with a scaling function of the distance between the points.} We call $\beta(x,\hat{p})^{-1}$ to the local temperature corresponding to the direction $\hat{p}$. High $\beta$ or, low temperature, corresponds to higher distinguishability. 

We argued in \cite{nosotros} that there is some degree of universality in these local temperatures and that the answer should be dictated essentially by the geometry. 
In order to compute these temperatures we have to study the local part of the modular Hamiltonian. 
 This is in a certain sense the opposite limit to the one pertaining to the first law of entanglement. This later gives $\Delta \langle {\cal H}_V\rangle=\Delta S$ for small deviation, low distinguishability,  between the states. Here we are interested in the limit of high distinguishability and high relative entropy.  

In the present paper we build on our previous work and study the local temperatures and the local part of the modular Hamiltonian in $d$ dimensions for free fields. We first write the general form of the spectrum of the modular Hamiltonian in a novel fashion in terms of solutions of the field equations in Euclidean space with some multiplicative boundary conditions in $V$. Then, we use an eikonal approximation to write these solutions. This captures the relevant physics of the problem and allows us to write the local terms of ${\cal H}$ for free scalar and fermion fields in a simple way. These are given in terms of the solutions of a purely geometric "eikonal" problem involving two orthogonal gradient fields $\vec{A}$ and $\vec{B}$ of the same modulus. In $d=2$ these become the Cauchy-Riemann equations for holomorphic functions.  We show the local inverse temperatures are given by the modulus of these vectors on the points of $V$. 

The eikonal problem, and the local temperatures, are the same for free scalar and fermions of any mass. 
As an example, we compute the local temperatures for a strip-like region, and find the explicit expression for $\beta(x,\hat{p})$. This is indeed dependent on direction. 

\section{Spectrum of the modular Hamiltonian for free fields}
We take a compact spatial region $V$ at $x^0=0$. We write the vacuum reduced density matrix as
\be
\rho = \mathrm{tr}_{\bar{V}} \ket{0}\bra{0}:=c \, \mathrm{e}^{-{\cal H}} \, , 
\ee
where ${\cal H}$ is the modular Hamiltonian and $\bar{V}$ is the complement of $V$. For free fields, ${\cal H}$ is quadratic in the fields, and then it is determined by certain numerical kernels (see for example the review \cite{review}). 

In order to understand the local structure of these modular Hamiltonians one needs a handle into the spectrum of these kernels. This essentially amounts to diagonalize the field correlator kernels inside $V$. This is an integral equation problem and it is  not easy to manage even in the high energy limit we are interested in this paper. For this reason we will find convenient to transform the kernel eigenvalue problem into another equivalent one which corresponds to find solutions of the field equations in $d$-dimensional Euclidean space with multiplicative boundary conditions on the $(d-1)$-dimensional region $V$. In this section, we explain this relation for scalar and fermion fields.  With this result in hand, in the next section we will be in position to use an eikonal approximation to the eigenvector problem in Euclidean space to find the structure of the local terms in the modular Hamiltonian.

\subsection{The scalar field} \label{section_scalar}

We will consider the theory of a free scalar field in $d$ dimensions. This field satisfies the Klein-Gordon equation and the equal time commutation relation $[\phi(\vec{x}),\pi(\vec{y})]=i\delta^{(d-1)}(\vec{x}-\vec{y})$ with $\pi(\vec{x}):=\partial_0 \phi(\vec{x})$.  The modular Hamiltonian for a spatial region $V$ is given as a quadratic expression \cite{review,peschel}
\be
{\cal H}=\frac{1}{2}\int_V d^{d-1}x\,d^{d-1}y\, \Big[ \pi(\vec{x}) N(\vec{x},\vec{y}) \pi(\vec{y}) + \phi(\vec{x}) M(\vec{x},\vec{y}) \phi(\vec{y}) \Big] \, . \label{smh}
\ee
The kernels are given in terms of the equal time correlators 
\bea
X(\vec{x}-\vec{y}):=\langle 0|\phi(\vec{x})\phi(\vec{y})|0\rangle \, , \\
P(\vec{x}-\vec{y}):=\langle 0|\pi(\vec{x})\pi(\vec{y})|0\rangle \, . \label{scorr}
\eea
Restricting the range of $x$ and $y$ to $V$, these define two integral operators in $V$, which we call for simplicity $X$ and $P$.   
Defining $C:=\sqrt{XP}$, we have \cite{review},
\bea
N &=& \frac{1}{C} \, \mathrm{log}\Bigg(\frac{C+\frac{1}{2}}{C-\frac{1}{2}}\Bigg) \, X \,, \label{N} \\
M &=& P \, \frac{1}{C} \, \mathrm{log}\Bigg(\frac{C+\frac{1}{2}}{C-\frac{1}{2}}\Bigg) \, .\label{M} 
\eea

In order to solve the spectrum of $X$ and $P$ as kernels in $V$, let us first consider a function $S(x)$ satisfying the Klein-Gordon equation in the Euclidean space $x:=(x_0,\vec{x}) \in \mathbb{R}^d$,
\be
(-\nabla^2+m^2)S(x)=0\,, \label{ekg}
\ee
everywhere except at the $(d-1)$-dimensional region $V$. On $V$, we impose the boundary conditions
\be
S^{+}(\vec{x}) =\lim_{x_0\rightarrow 0^+} S(x_0,\vec{x})=\lambda \lim_{x_0\rightarrow 0^-} S(x_0,\vec{x})=\lambda \, S^-(\vec{x}) \hspace{0.5 cm} \forall \vec{x}\in V \, .  \label{bcs} 
\ee
Further, we require the regularity condition at infinity
\be
\lim_{|x|\rightarrow \infty} S(x)=0 \, ,
\ee
and demand that $|S|$ remains bounded  at the boundary of the region $V$. This last condition gives the right uniqueness class of the solutions (see Appendix \ref{ap1}) to treat the density matrix of the vacuum state \cite{review}. Additional solutions exist with divergences for $x\rightarrow \partial V$, but these are related to problems where a magnetic vortex is placed on the boundary.  

We define $\mathcal{M}:=\mathbb{R}^d - V$ for the region in the Euclidean plane where (\ref{ekg}) is satisfied. Consider now the Euclidean Green function $G_S$ for the scalar field in the full Euclidean space without the cut at $V$,\footnote{$K_{\nu}$ is the modified Bessel function of second kind of order $\nu$.}
\be
 G_s(x-y)= \frac{1}{(2\pi)^d} \int_{\mathbb{R}^d} \frac{1}{p^2 + m^2} \mathrm{e}^{i p \cdot (x-y)} = \frac{m^{d-2}}{(2 \pi)^{d/2}} (m|x-y|)^{1-\frac{d}{2}} \, K_{1-\frac{d}{2}}(m|x-y|) \, , \label{gs}
\ee
where $p \cdot x= p_\mu x_\mu$ is the Euclidean scalar product. The scalar Green function (\ref{gs}) satisfies the inhomogeneous equation
\be
(-\nabla_x^2+m^2) G_S(x-y)= \delta^{(d)}(x-y) \, . \label{egs}
\ee
Combining (\ref{ekg}) and (\ref{egs}) we have the current
\be
J_{\mu}^{x}(y)=\partial^y_\mu G_s(x-y) S(y)-G_s(x-y) \partial_\mu S(y) \, , \label{current}
\ee
which is conserved everywhere in $\mathcal{M}$ except at the coincident points $x=y$ where,
\be
\partial_\mu J_\mu^{x}(y)= -\delta^{(d)}(x-y) S(y) \, . 
\ee
Integrating this equation on $\mathcal{M}$ we obtain from Stokes theorem
\bea
-S(x) &=& \int_\mathcal{M} d^d y \, \partial_\mu J_\mu^{x}(y)  \nonumber \\
 &=& -\int_V d^{d-1}y \, \bigg(\partial^y_0 G_s(x-(0,\vec{y}) ) \, S^+(\vec{y}) - G_s(x-(0,\vec{y}) ) \,\partial_0 S^+(\vec{y})\bigg)  \nonumber \\ 
&& + \int_V d^{d-1}y \, \bigg(\partial^y_0 G_s(x-(0,\vec{y}) ) \,S^-(\vec{y}) - G_s(x-(0,\vec{y}) ) \, \partial_0 S^-(\vec{y})\bigg) \,, \label{ints}
\eea
where we drop the boundary contribution at infinity since the Green function (\ref{gs}) vanishes exponentially at infinity. Using the boundary condition (\ref{bcs}) we simplify (\ref{ints}) to get 
\be
S(x) = (1- \lambda^{-1}) \int_V d^{d-1}y \, \bigg(\partial^y_0 G_s(x-(0,\vec{y}) ) \, S^+(\vec{y}) - G_s(\vec{x}-(0,\vec{y}) )\, \partial_0 S^+(\vec{y})\bigg) \, , \label{ints2}
\ee
which gives us the function $S(x)$ on the Euclidean plane from the values of the function and its time derivative on the region $V$. Now, we are going to take  the limit of (\ref{ints2}) when $x$ goes to $V$ from above, that is $x\rightarrow (0^+,\vec{x}) \in V$ with $\vec{y}\in V$. We use 
\bea
\lim_{x\rightarrow V^+ ,\,y\in V} G_s(x-y)= X(\vec{x} - \vec{y}) \, ,  \label{limpx1}\\
\lim_{x\rightarrow V^+ , \,y\in V} \partial_0^y G_s(x-y)= \frac{1}{2}\delta^{(d-1)}(\vec{x} - \vec{y}) \, , \label{limpx2} \\
  \lim_{x\rightarrow V^+ , \,y\in V} -\partial_0^x \partial_0^y G_s(x-y)= P(\vec{x} - \vec{y}) \,.\label{limpx3} 
\eea

Calling 
\bea
u_\lambda(\vec{x})&:=&S^+(\vec{x})\,,  \label{mira1}\\
 v_\lambda(\vec{x})&:=&\partial_0 S^+(\vec{x})\label{mira2}\,,
\eea 
  and replacing (\ref{limpx1}-\ref{limpx2}) in (\ref{ints2}), we get
\be
\int_V d^{d-1}y \, X(\vec{x}-\vec{y}) v_\lambda(\vec{y})= \frac{1+\lambda}{2(1-\lambda)} u_\lambda(\vec{x}) \, . \label{eigenX}
\ee
 An analogous equation follows from taking a $\partial^x_{0}$ derivative in (\ref{ints2}) before taking the limit,\footnote{Notice the kernel $P$ is singular for $x\rightarrow y$ and its distributional definition is given by eq. (\ref{limpx3}). }
\be
\int_V d^{d-1}y \, P(\vec{x}-\vec{y})  u_\lambda(\vec{y})= \frac{1+\lambda}{2(1-\lambda)} v_\lambda(\vec{y})\,. \label{eigenP}
\ee
 From here $v_\lambda(\vec{x})$ and $u_\lambda(\vec{x})$ are eigenvectors of $PX$ and $XP$ respectively. We have in short notation
\bea
PX \,v_\lambda= \frac{1}{4} \left(\frac{1+\lambda}{1-\lambda}\right)^2\, v_\lambda \, ,\label{refuno} \\
XP\, u_\lambda=  \frac{1}{4} \left(\frac{1+\lambda}{1-\lambda}\right)^2\, u_\lambda \, . \label{refdos}
\eea

Since the product $XP$ has eigenvalues in $(1/4,\infty)$ \cite{review}, it follows that $\lambda >0$. 
However, $\lambda$ and $\lambda^{-1}$ give place to the same eigenvectors, except for a global sign change on $v_\lambda$. This is due to the time reflection symmetry of the problem.    
Hence, we can restrict to $\lambda\in (0,1)$. We write this introducing a new parameter $s$ as
\be
\lambda = \mathrm{e}^{-2 \pi s} \hspace{0.5 cm} s \in \mathbb{R^+} \, . \label{pin}
\ee
Owing to possible degeneracies parametrized by an index $k$, from now we will name the solutions as $u_s^k$ and $v_s^k$. Moreover, as shown in the Appendix \ref{ap1}, the solutions are uniquely determined by the asymptotic behaviour at the vicinity of the boundary $\partial V$. We expect $d-2$ degeneracy parameters, and on top of that, we expect some finite labels if $\partial V $ has more than one connected components.    

Since $XP$ and $PX$ are not self-adjoint operators, we have that the basis\footnote{$XP$ has continuum spectrum and $\{u_s^k\}$ and $\{v_s^k\}$ are basis in the generalized sense.}  {$\{u_s^k \}$ and $\{v_s^k \}$ are not orthonormal. However, these basis can be chosen to be dual to each other, since, due to (\ref{refuno}) and (\ref{refdos}), 
\be
\langle v_{s'}^{k'}| XP|u_s^k\rangle= \frac{\coth(\pi s )^2}{4} \langle v_{s'}^{k'}|u_s^k\rangle=\frac{\coth(\pi s' )^2}{4} \langle v_{s'}^{k'}|u_s^k\rangle\,,
\ee
and in consequence, the vectors $u_s^k$ and $v_{s'}^{k'}$ are orthogonal for $s\neq s'$. We then choose the normalization such that
\be
\langle v_{s'}^{k'}|u_s^k\rangle=\int_V d^{d-1}x\, v_{s'}^{k'}(\vec{x})^* u_s^k(\vec{x})=\delta(s-s')\delta(k-k')\,,
\ee
where the $\delta(k-k')$ over the variables $k$, $k'$ may include a discrete part.

Using the solutions to the "eigenvalues" problems (\ref{eigenX}) and (\ref{eigenP}), we can easily decompose the operators $X$, $C$ and $P$ as
\bea
X(\vec{x}-\vec{y}) &=& \frac{1}{2}\int dk \,\int_{\mathbb{R^+}} ds \, u_s^k(\vec{x}) \,  \mathrm{coth}(\pi s) \, u_s^k(\vec{y})^* \,, \\
P(\vec{x}-\vec{y}) &=&\frac{1}{2}\int dk \, \int_{\mathbb{R^+}} ds \, v_s^k(\vec{x}) \,  \mathrm{coth}(\pi s) \, v_s^k(\vec{y})^*  \,,\\
C(\vec{x}-\vec{y}) &=& \frac{1}{2}\int dk \,\int_{\mathbb{R^+}} ds \, u_s^k(\vec{x}) \,  \mathrm{coth}(\pi |s|) \, v_s^k(\vec{y})^* \, .  
\eea

This is the desired result relating the spectral decomposition of the kernels $X$ and $P$ in the $(d-1)$-dimensional region $V$ with the solutions of the $d$-dimensional Klein-Gordon equation through (\ref{mira1}-\ref{mira2}). Notice the eigenvalues of the kernels are mapped to the factor in the boundary condition on $V$  (eq. (\ref{bcs})).     

Replacing these formulae into (\ref{N}) and (\ref{M}), we finally obtain the following expressions for the modular Hamiltonian kernels $N$ and $M$
\bea
N(\vec{x},\vec{y}) &=& \int dk \, \int_{\mathbb{R^+}} ds \, u_s^k(\vec{x}) \, 2 \pi s \, u_s^k(\vec{y})^* \,, \label{MN1} \\
M(\vec{x},\vec{y}) &=& \int dk \, \int_{\mathbb{R^+}} ds \, v_s^k(\vec{x}) \, 2 \pi s \, v_s^k(\vec{y})^* \, . \label{MN2} 
\eea

To summarize, in this section we have shown that solving the Klein-Gordon equation in Euclidean space with the multiplicative boundary condition (\ref{bcs}) we can construct the eigenfunctions $u_s^k$ and $v_s^k$ which diagonalize the correlators (\ref{scorr}) as kernels in the region $V$. These eigenfunctions are related to boundary values of the Klein-Gordon solution and its derivative on the cut. Using that, we can easily obtain the modular Hamiltonian (\ref{smh}) in term of these eigenfunctions as we expressed in (\ref{MN1}-\ref{MN2}).

\subsection{The Dirac field}

In this section, we will apply to the Dirac field the same idea used above for the case of a scalar field. This field satisfies the Dirac equation and the equal time anti-commutation relations $\{\psi(\vec{x}),\psi^\dagger(\vec{y})\}=\delta^{(d-1)}(\vec{x}-\vec{y})\,\boldsymbol{1}_{n\times n}$ where $n=2^{\lfloor \frac{d}{2} \rfloor}$. Now, the modular Hamiltonian can be written as
\be
{\cal H}=\int_V d^{d-1}x\,d^{d-1}y\, \psi^\dagger(\vec{x})  H (\vec{x},\vec{y}) \psi(\vec{y}) \, , \label{modH_dirac}
\ee
where the Hamiltonian kernel $H(\vec{x},\vec{y})$ can be expressed in terms of the correlator matrix kernel 
\be
C(\vec{x}-\vec{y})= \langle 0|\psi(\vec{x})\psi^\dagger(\vec{y})|0 \rangle 
\ee
as \cite{peschel}
\be
H= - \log (C^{-1} - 1) \, . \label{dhk}
\ee

We now consider the Dirac equation in Euclidean space.\footnote{Here we are using the Euclidean version of the gamma matrices which satisfies $\{\gamma_\mu ,  \gamma_\nu\} = 2 \delta_{\mu \nu} \boldsymbol{1}_{n\times n}$ and are related to the Minkowski gamma matrices as $ \gamma_E^{0} = \gamma_M^{0}$ and $ \gamma_E^{k} = - i \gamma_M^{k}$ for $k=1, \ldots , d-1$ where we use the convention $ \gamma_{E \, \mu} = \gamma_E^{\mu}$}  The spinor field $S(x)$ satisfies
\be
(\gamma_\mu \partial_\mu +m) S(x)=0 \label{edir}
\ee
everywhere except at the region $V$, where we also impose the same multiplicative boundary conditions with factor $\lambda$ used for the scalar field (\ref{bcs}). The solution must vanish at infinity and, in addition, $l^{1/2} S(x)$, where $l$ is the distance from $x$ to the boundary of the region $\partial V$, has to remain bounded as $x\rightarrow \partial V$ \cite{review}.  

We also take the Euclidean Green function $G_D$ for the Dirac field in the full space without the cut,
\be
 G_D(x-y) = (- \gamma_\mu \partial^x_\mu +m) G_S (x-y) = \frac{1}{(2\pi)^d} \int_{\mathbb{R}^d} \frac{-i \gamma_\mu p_\mu + m}{p^2 + m^2} \mathrm{e}^{i p \cdot (x-y)}  \, . \label{gd}
\ee
 The Dirac Green function (\ref{gd}) satisfies the relation $G_D(\xi) = G_D^\dagger(\xi)$ and the inhomogeneous equation
\be
(\gamma_\mu \partial^x_\mu +m) G_D(x-y)= \delta^{(d)}(x-y) \, . \label{egd}
\ee
Combining (\ref{edir}) and (\ref{egd}) we have the current
\be
J^{x}_\mu(y)= G_D (x-y)\gamma_\mu S(y) \, , 
\ee
which is conserved everywhere in $\mathcal{M}=\mathbb{R}^d - V$ except at the points $y=x$ where,
\be
\partial_\mu J_\mu^{x}(y)= -\delta^{(d)}(x-y) S(y) \, . 
\ee
Integrating this equation on $\mathcal{M}$ we obtain from Stokes theorem
\be
-S(x) = \int_\mathcal{M} d^d y \, \partial_\mu J_\mu^{x}(y) = -\int_V d^{d-1}y \, G_D (x-(0,\vec{y}))\gamma_0 S^+(\vec{y}) + \int_V d^{d-1}y \,  G_D (x-(0,\vec{y}))\gamma_0 S^-(\vec{y}) \,, \label{intd}
\ee
where we drop the boundary contribution at infinity since the Green function (\ref{gs}) vanishes exponentially at infinity. Using the boundary condition (\ref{bcs}) we simplify (\ref{intd}) into
\be
S(x) =  (1 - \lambda^{-1}) \int_V d^{d-1}y \, G_D (x-(0,\vec{y}))\gamma_0 S^+(\vec{y})  \,, \label{intd2}
\ee
which gives us the function $S(x)$ on the Euclidean plane from its values on $V$. 

Taking the limit $x \rightarrow (0^+,\vec{x}) \in V$ with $\vec{y} \in V$, and using 
\be
\lim_{x \rightarrow V^+,\, y\in V} G_D(x-y)\gamma_0= C(\vec{x} - \vec{y}) \, 
\ee
we get
\be
\int_V d^{d-1}y\, C(\vec{x}-\vec{y}) S^+(\vec{y})=\frac{\lambda}{\lambda-1} S^+(\vec{x}) \, .
\ee
That is, the boundary value of the spinor $S(x)$ plays the role of eigenvector of the correlator kernel on $V$.
Since eigenvalues of $C(\vec{x} - \vec{y})$  are restricted to $(0,1)$ (see \cite{review}) the range of possible $\lambda$ is $\lambda \in (-\infty,0)$. Then, the allowed range for the boundary condition factor is the negative numbers, in contrast to the scalar case for which $\lambda >0$.  We can write equivalently
\be
\lambda = - \mathrm{e}^{-2 \pi s} \hspace{0.5 cm} s \in \mathbb{R} \, . 
\ee
Notice $s$ and $-s$ now give two independent eigenvectors of the kernel $C(x-y)$. They correspond to time-reflected solutions of the Euclidean problem,
\be
S_{-s}(x_0,\vec{x})=M \,S^*_s(-x_0,\vec{x})\,,
\ee
where $M$ is the matrix given time inversion symmetry, determined by the relations $M\gamma_0^{*}=-\gamma_0 M$, $M\gamma_i^*=\gamma_i M$.

We name the eigenvectors as 
\be
u_s^k(x) := S_s^{+\,k}(x)\,,\hspace{.7cm}x\in V\,,
\ee
where $k$ identifies the different solutions with the same $s$. 
 
Since, in this case, the kernel $C$ is a self-adjoint operator\footnote{By the hemiticity condition of the Wightman distributions we have that $C_{ij}(\vec{x}-\vec{y})=C_{ji}(\vec{y}-\vec{x})^*$.} we can choose the eigenfunctions $\{u_s^k\}$ to be orthonormal
\be
\langle u_s^k | u_{s'}^{k'}\rangle= \int_V d^{d-1}x \, u_s^{k\,\dagger}(\vec{x}) \, u_{s'}^{k'}(\vec{x}) = \delta(s-s^\prime)\delta(k-k') \,.\label{choose}
\ee
The correlator kernel writes
\be
C(\vec{x}-\vec{y}) = \int dk\, \int_{\mathbb{R}} ds \, u_s^k(\vec{x}) \, \frac{1}{1+\mathrm{e}^{2 \pi s}} \, u_s^{k\, \dagger}(\vec{y}) \, . 
\ee
Finally, using this formula and (\ref{dhk}) we obtain the following expression for the Hamiltonian kernel $H$
\be
H(\vec{x},\vec{y}) = -\int dk\, \int_{\mathbb{R}} ds \, u_s^k(\vec{x}) \, 2 \pi s \, u_s^{k \dagger}(\vec{y}) \, . \label{H_ker} 
\ee

\section{Eikonal approximation and local temperatures}
In general it is a difficult problem to find solutions of the wave equation with the specified boundary conditions on $V$. However, here we are interested in the eigenvectors of the modular Hamiltonian kernels for the sake of understanding the local terms around a point, that is, how these kernels behave in the limit $x\sim y$. This allows us to simplify the problem in the following terms. First we can see the local terms are naturally related to the high frequency limit of the solutions. This in turn implies we have to look at the spectrum for the limit of large multiplicative factor, $|s|\gg 1$. A simple exact example can illustrate this fact. Consider the case of a massless Dirac field in $d=2$ for $n$ intervals. The solutions are of the form \cite{diracd2}
\be
u^k_s(x_1)= g^k(x_1) e^{-i s w(x_1)}\,,  
\ee  
where $g^k(x_1)$ is some smooth, non oscillatory pre-factor, and 
\be
w(x_1)=\log\left(-\prod_{i=1}^n\frac{x_1-l_i}{x_1-r_i}\right)\,. \label{funi}
\ee  
Here $(l_i,r_i)$, $i=1,\cdots,n$ are the different non-intersecting intervals. It is evident that, when these solutions are inserted into (\ref{H_ker}), and we look at the $x\sim y$ limit of the kernel, we will need only the large $s$ limit in the Fourier integral. The result will depend essentially on the function $w(x)$, since the pre-factors are fixed to give the correct normalization for the local ``plane wave'' type contribution. We get that the local temperatures are given exclusively in terms of the phase factor by \cite{diracd2}
\be
\beta(x_1)=2 \pi w'(x_1)^{-1}=2 \pi \left(\sum_{i=1}^n\left(\frac{1}{x_1-l_i}+\frac{1}{r_i-x_1}   \right)\right)^{-1}\,.\label{timeo}
\ee 

As argued in \cite{nosotros}, the limit of large $s$ can be understood as a limit of large angular momentum for the angular variable describing the transition from one side of the cut to the other. Hence, this limit is one of large gradients for the solutions $S$, and we can use an eikonal approximation. In this approximation the mass of the field becomes irrelevant since it is much smaller than the typical kinetic energies involved \cite{nosotros}. 

This eikonal limit will allows us to extract  a universal geometric prescription for the local temperatures. Notice that even if we are making an approximation for the solutions, this approximation becomes exact for the large $s$ limit, and the result for the local temperatures will then be exact, since these are defined for the limit of high energy-small size excitations.

\subsection{Scalar field}

Therefore we want to solve the Euclidean Klein-Gordon equation (\ref{ekg}) for large $s$. In the spirit of the eikonal approximation we write 
\be
S(x)=g(x) \, \mathrm{e}^{\alpha(x)} \, , \label{sol_eik}
\ee
where $\alpha \sim \mathcal{O}(s)$ but $g(x)$ is a normalization factor of polynomial order in $s$. Applying the equation of motion (\ref{ekg}) to this parametrization of the solution we get
\be
0= (-\nabla^2+m^2)S(x)= - \mathrm{e}^{\alpha} [(\nabla \alpha)^2 \, g+ \nabla^2 \alpha \, g + 2 (\nabla \alpha) \, (\nabla g) +(\nabla g)^2  - m^2 \, g]  \, . \label{eik_ekg}
\ee
 Keeping only the leading terms for large $s$ we get
\be
(\nabla \alpha)^2 = 0  \, . \label{eikonal_scalar}
\ee
Since the function $\alpha$ could be (and must be) complex valued, we have non trivial solutions for this last equations. Writing 
\be 
\alpha= a+ i b
\ee
 with $a$ and $b$ real valued functions we can rewrite (\ref{eikonal_scalar}) as
\bea
(\nabla a)^2 = (\nabla b)^2\,,   \\
\nabla a  \cdot \nabla b = 0 \, ,   \label{eikonal_scalar_2}
\eea
with boundary conditions (see (\ref{bcs}))
\bea
&& a(0^+,\vec{x}) = a(0^-,\vec{x}) - 2 \pi s\,, \hspace{0.7 cm} \forall \vec{x}\in V \,, \label{eik_bb}\\
&& b(0^+,\vec{x}) = b(0^-,\vec{x}) \, . \label{eik_bc}
\eea
We see $a$ and $b$ are defined up to an additive constant. Given a solution $\alpha=a+i b$ of these equations the complex conjugate $\alpha^*=a -i b$ also gives a solution for the same value of $s$.   Likewise, changing $a(x_0,\vec{x})\rightarrow -a(-x_0,\vec{x})$, $b(x_0,\vec{x})\rightarrow b(-x_0,\vec{x})$ also gives a solution of the same problem. This leaves the asymptotic behaviour of $\alpha$ near $\partial V$ unchanged. Then they represent approximate solutions for $S$ with the same asymptotic behaviour and by uniqueness (see the discussion around eq. (\ref{v}) in the Appendix \ref{ap1}) we expect they represent the same solution of the eikonal problem, up to an additive constant, 
\be
-a(-x_0,\vec{x})=a(x_0,\vec{x})+\textrm{const}\,,\hspace{.7cm} b(-x_0,\vec{x})=b(x_0,\vec{x})+\textrm{const}\,.\label{imi}
\ee
This implies in particular that the gradient $\nabla a$ is orthogonal to $V$ and $\nabla b$ is parallel to $V$. Without loss of generality we can fix $a(0^+,\vec{x})=0$, $a(0^-,\vec{x})=2\pi s$, for $\vec{x}\in V$.    

Factoring out a linear $s$ dependence, this system can be written in terms of two vector fields $A(x)=s^{-1}\nabla a(x)$ and $B(x)=s^{-1}\nabla b(x)$ in $\mathbb{R}^d$, that are continuous outside $\partial V$, and obey
\bea
| A | &=& | B |  \,, \label{5}\\
A  \cdot B &=& 0 \, ,  \label{eik_scalar}\\
 \partial_i A_j-\partial_j A_i &=& 2 \pi  \,  (\xi^1_i \xi^2_j -\xi^2_i \xi^1_j)\, \delta^{d-2}_{\partial V}\,, \label{trtr}\\
\partial_i B_j-\partial_j B_i &=& 0 \, ,\label{6}  
\eea
where $\xi^1=(1,0,\cdots,0)$,  and $\xi^2$ is a unit outward-pointing vector normal to $\partial V$ and $\xi^1$,  and $\delta^{d-2}_{\partial V}$ is the delta function on $\partial V$.\footnote{This is defined such that $\int d^{d}x\, f(x) \delta^{d-2}_{\partial V}(x)=\int_{\partial V} dy_\parallel\, \sqrt{g(y_\parallel)} f(y_\parallel)$, for $y_\parallel$ coordinates on $\partial V$. } 
Eq. (\ref{trtr}) just means that the circulation of $A$ around $\partial V$ in the positive time direction is $2\pi$.  
The previous discussion (\ref{imi}) also implies that, writing $T$ for the time inversion matrix in $\mathbb{R}^d$, 
\bea
A(x)&=&-T A(T x)\,,\label{7}\\
B(x)&=& T B(T x)\,.\label{8}
\eea

These are the eikonal equations. Geometrically, $a$ and $b$ are two orthogonal coordinates in $\mathbb{R}^d$, and $a$ is an angular coordinate that goes from $0$ to $2\pi s$ between the two sides of the cut. In addition, the gradients of the two coordinates have equal modulus.    

\subsubsection{Eigenfunctions and normalization}
Suppose that we have solutions $(A_k , B_k)$ of this system parametrized with the multi-index $k=(k_1, \ldots , k_{d-2})$, in some domain $k\in {\cal K}$. We expect in $d$ dimensions, a degeneracy in the eigenspace of solutions which could be labeled with $d-2$ parameters $k_i$. In general, according to the discussion in the Appendix \ref{ap1}, we expect these parameters to label momentum variables associated to the description of $\partial V$; they will be continuous for unbounded boundaries and discrete for bounded ones, additional discrete labels may occur for multiple component regions. For simplicity of notation, in what follows we will work with continuum variables $k_i$.

In terms of these functions, we can write
\be
S_k(x) = g_k(x) \, \mathrm{e}^{s \int_{x_*}^{x} \, A_k(y) \cdot dy} \mathrm{e}^{i s \int_{x_*}^{x} \, B_k(y) \cdot dy} \, , \label{eik_sol}
\ee
where $x_*$ is a fixed arbitrary point and the integrals on the exponent are line integrals of any path which connects $x_*$ to $x$.\footnote{The integral is independent of the chosen path since $A$, $B$ are gradients outside $\partial V$.} For convenience we set $x_*\in V$. From this, and considering that $\vec{A}=0$ for $x\in V$,  we get the eigenfunctions of the kernels on $V$ as
\bea
u_s^k(\vec{x}) &=& g_k(\vec{x}) \,  \mathrm{e}^{i s \int_{x_*}^{\vec{x}} \, \vec{B}_k(\vec{y}) \cdot d\vec{y}}  \,, \\
v_s^k(\vec{x}) &=& g_k(\vec{x}) \,s \,  A_0^k (\vec{x}) \mathrm{e}^{i s \int_{x_*}^{\vec{x}} \, \vec{B}_k(\vec{y}) \cdot d\vec{y}} \, ,
\eea
where we have written $A_0^k(\vec{x})$ for the time component (the only non zero component) of the vector $A_k(x)$ for $x\in V$.

The scalar product of the eigenfunctions
\be
\langle u_{s'}^{k'} | v_s^{k}\rangle= \int_V d^{d-1}x \,s\,  g_k(\vec{x}) g^*_{k'}(\vec{x}) \, A_0^k(\vec{x}) \, \mathrm{e}^{-i s' \int_{x_*}^{\vec{x}} \, \vec{B}_{k'}(\vec{y}) \cdot d\vec{y} +i  s \int_{x_*}^{\vec{x}} \, \vec{B}_{k}(\vec{y}) \cdot d\vec{y} }  \,, 
\ee
should be normalized to give
\be 
\delta(s-s') \delta^{(d-2)}(k-k')= \delta(s-s')\delta(k_1-k_1') \ldots \delta(k_{d-2} -k_{d-2}')\,.
\label{delt}\ee
This would be the case for the exact solutions; in the eikonal approximation, due to the large exponents involved and in order to get the right coefficient of delta function, we can approximate the exponential factor with its Taylor series on  $\Delta k_i=k_i'-k_i$,
\be
\langle u^{k'}_{s'} | v^k_s \rangle= \int_V d^{d-1}x \, g_k(\vec{x}) g^*_{k}(\vec{x}) \,s\, A_0^k(\vec{x}) \mathrm{e}^{-i \Delta s\, (\int_{x_*}^{\vec{x}} \, \vec{B}_k(\vec{y}) \cdot d\vec{y}) }\mathrm{e}^{-i s \,\Delta k_i\, \partial_{k_i} \,(\int_{x_*}^{\vec{x}} \, \vec{B}_k(\vec{y}) \cdot d\vec{y}) } \, . \label{ortho_scalar}
\ee

We define $d-1$ functions
\be
 \sigma_i(\vec{x}) = \partial_{k_i} \Bigg(\int_{x_*}^{\vec{x}} \, \vec{B}_k(\vec{y}) \cdot d\vec{y} \Bigg)  \hspace{0.5 cm}  i \in \{1, \ldots , d-2\}\,,\hspace{.5cm} \sigma_{d-1}(\vec{x})= \int_{x_*}^{\vec{x}} \, \vec{B}_k(\vec{y}) \cdot d\vec{y}\, .
\ee
which are $\mathcal{O}(1)$ in the eikonal parameter $s$. We assume we can invert the above functions in the region $W\subseteq V$ where the eikonal functions are non vanishing, 
\be
\vec{x}=(x_1,\ldots , x_{d-1}) \in W \hspace{0.2 cm} \leftrightarrow \hspace{0.2 cm} (\sigma_1, \ldots , \sigma_{d-1})\in \Sigma \, , \label{cv_scalar}
\ee
where $\Sigma \in \mathbb{R}^{d-1}$  is the domain where these new variables are defined (thus, the range of the functions $\sigma_i$). After making the change of variables (\ref{cv_scalar}), the formula (\ref{ortho_scalar}) becomes
\be
\langle u_{s'}^{k'} | v_{s}^k\rangle= \int_\Sigma d^{d-1}\sigma \, g_k g^*_k \,s\, A_0^k \, J^{-1} \, \mathrm{e}^{-i \Delta s \, \sigma_{d-1} }\, \mathrm{e}^{-i\, s \,\sigma_i   \Delta k_i } \, , 
\ee
where $J$ is the Jacobian determinant of the change of variables matrix (\ref{cv_scalar})
\be
J = \left| \det\left(\frac{\partial ( \sigma_1,\cdots,\sigma_{d-1})} { \partial (x_1,\cdots,x_{d-1})}\right)\right| \, , \label{J_scalar}
\ee
and we have the relations 
\be
\partial_{x_j}\sigma_i = \partial_{k_i} B_j\,, \,i=1,\cdots,d-2\,,\hspace{.7cm}\partial_{x_j}\sigma_{d-1}=B_j\,.
\label{dio}
\ee

The fact that the functions $u_s^k$, $v_s^k$, are orthogonal, shows that the pre-factors $g_k$ satisfy 
\be 
|g_k|^2 = \frac{s^{d-3}\, J}{(2 \pi)^{d-1} \, A_0^k}\,,
\ee
where the overall normalization has been chosen such that (\ref{choose}) holds. Notice $A_0^k$ should be always positive on $V$.

Then, except for a point dependent ${\cal O}(1)$ phase which is not relevant for the applications we have in mind, in the eikonal approximation, the appropriately normalized functions $u_s^k$ and $v_s^k $ can be chosen as 
\bea
u_s^k(\vec{x}) = \sqrt{\frac{s^{d-3}\, J(\vec{x})}{(2 \pi)^{d-1}\, A_0^k(\vec{x})}} \,  \mathrm{e}^{i \,s \,\int_{x_*}^{\vec{x}} \, \vec{B}_k(y) \cdot dy}  \,, \\
v_s^k(\vec{x}) = \sqrt{\frac{s^{d-1}\, J(\vec{x})\,   A_0^k(\vec{x})}{(2 \pi)^{d-1}}} \; \mathrm{e}^{i \,s \,\int_{x_*}^{\vec{x}} \, \vec{B}_k(y) \cdot dy} \, .
\eea

\subsubsection{Local terms in the modular Hamiltonian}
Now, using the formulae (\ref{MN1}-\ref{MN2}) we calculate the kernels $N$ and $M$ of the modular Hamiltonian in the local limit. We write the two variables of the kernels as $\vec{x}',\vec{y}'$, where we look at the limit $\vec{x}'\sim \vec{y}'$ and both these variables are in a small neighborhood of a point $\vec{x}$ that we take fixed. For $N$, we have
\bea
N(\vec{x}',\vec{y}') &\simeq& \int_0^\infty ds\,\int_{\cal K} d^{d-2}k \,\, \pi\, s  \, \sqrt{\frac{s^{d-3}\,J(\vec{x}')}{(2 \pi)^{d-1}\, A_0^k(\vec{x}')}} \sqrt{\frac{s^{d-3}\,J(\vec{y}')}{(2 \pi)^{d-1} \,A_0^k(\vec{y}')}} \, \mathrm{e}^{i s\,\int_{x_*}^{\vec{x}'} \, \vec{B}_k(\vec{y}) \cdot d\vec{y} - i s\,\int_{x_*}^{\vec{y}'} \, \vec{B}_k(\vec{y}) \cdot d\vec{y}} \, , \nn \\
&\simeq & \int ds\,\int_{\cal K} d^{d-2}k \,  \pi  \, \frac{s^{d-2}\, J(\vec{x})}{(2 \pi)^{d-1} A_0^k(\vec{x})}  \, \mathrm{e}^{i \,s\,\vec{B}_k(\vec{x}) \cdot \Delta\vec{x}'} \,,
\eea
where in the second line we set  $ \Delta \vec{x}'= \vec{x}'-\vec{y}'$, with $|\Delta \vec{x}' |$ small, and neglected factors of $\Delta \vec{x}'$ everywhere except in the exponent, since we are looking for the leading local term. 

Next, keeping in mind we are looking at a neighborhood of a point $x$, we make a change of variables (at fixed $\vec{x}$)  
\be
(s,k_1,\ldots , k_{d-2}) \hspace{0.1 cm} \leftrightarrow \hspace{0.1 cm} \vec{p}=s\,(B_1, \ldots, B_{d-1})\,,\label{changee}
\ee
 whose Jacobian determinant is $\Big| \frac{\partial(s B)}{\partial(s,k)}\Big|_{\vec{x}} \equiv s^{d-2} J(\vec{x}) $, i.e., because of the relations (\ref{dio}), it is proportional to the expression (\ref{J_scalar}). Then, for the local term of the kernel $N$ we have
\be
N_{\textrm{loc}}(\vec{x}',\vec{y}') = \int \frac{d^{d-1}p}{(2\pi)^{d-1}}\,   \, \frac{\pi}{A_0(\hat{p},\vec{x})} \, \mathrm{e}^{i \vec{p} \cdot (\vec{x}'-\vec{y}')} \,. \label{2}
\ee
Note the momentum variable of the local part of the kernel is played essentially by the solution $B$ at the point, $\vec{p}=s\,\vec{B}(\vec{x})$. The time component $A_0^k(\vec{x})$ will depend on the particular solution of the eikonal equation that gives the direction $\hat{p}=\hat{B}(\vec{x})$. This dependence arises because the $k$ dependence has been trade off to a $\vec{B}(\vec{x})$ dependence in the change of variables. This is why we write
\be
A_0(\hat{p},\vec{x}):= A_0^{k(\hat{p},\vec{x})}(\vec{x})\,.
\ee
 Notice that $A_0=|A|=|\vec{B}|$ on the points of $V$ (recall also  (\ref{eik_scalar})). Hence $\pi/A_0(\hat{p},\vec{x})$ can also be written as $\pi s/|\vec{p}|$, where now $s$ has to be understood as a function of $\hat{p}$ and $\vec{x}$.    

Following the same steps for the kernel $M$ we obtain
\be
M_{\textrm{loc}}(\vec{x}',\vec{y}') = \int \frac{d^{d-1}p}{(2\pi)^{d-1}}   \,\frac{\pi\, \vec{p}^2 }{A_0(\hat{p},\vec{x})} \, \mathrm{e}^{i \vec{p} \cdot (\vec{x}'-\vec{y}')} \, . \label{3}
\ee

Expressions (\ref{2}) and (\ref{3}) have exactly the expected form for a scalar \cite{nosotros}. In coordinate space the expressions are less transparent. $N_{\textrm{loc}}$ is a homogeneous distribution of degree $d-1$ which can include a component of the delta function $\delta^{d-1}(\vec{x}'-\vec{y}')$ but more generally contains angle dependent terms of the form \cite{nosotros}
\be
\frac{f\left(\frac{\vec{x}'-\vec{y}'}{|\vec{x}'-\vec{y}'|}\right)}{|\vec{x}'-\vec{y}'|^{d-1}}\,.
\ee
 $M_{\textrm{loc}}$ is a homogeneous distribution of degree $d+1$ and may contain second derivatives of the delta function as in the local energy density operator, but more generally it also contains angle dependent terms, with the same scaling dimensions $d+1$. 

To find the local temperatures in terms of $A_0(\hat{p},\vec{x})$ we compute the operator form of the local Hamiltonian by inserting these kernels into (\ref{smh}). Of course, we only know the leading local structure of the kernels around a point $\vec{x}$, and integrating it with  the field operators give us an operator expression that is only valid as leading term for high energy localized excitations. In accordance with this, we are using the massless field expression
\be
\phi(x)=\int \frac{d^{d-1} p}{(2\pi)^{\frac{d-1}{2}}\sqrt{2|\vec{p}|}}\, \left(a_{\vec{p}}\, e^{-i p x}+ a^\dagger_{\vec{p}} \,e^{i p x}  \right)\,.
\ee
Computing the modular Hamiltonian (\ref{smh}) with $M$ and $N$ as they were given by (\ref{2}) and (\ref{3}) with $x',y'$ in all space, we get
\be
{\cal H}_{\textrm{loc}}=\int d^{d-1}p\, \left(\frac{\pi}{A_0(\hat{p},\vec{x})}+\frac{\pi}{A_0(-\hat{p},\vec{x})}\right)  \, |\vec{p}|\, a^\dagger_{\vec{p}}a_{\vec{p}} \,.
\ee
 $A_0(\hat{p},\vec{x})$ is in fact $A_0(\hat{B},\vec{x})$, that is, the value of $A_0=|B|$ on a point $\vec{x}$ for a solution of the eikonal equations with $B$ at the point $\vec{x}$ pointing in the direction of $\hat{p}$. We recall that the eikonal equations having a solution $(A,B)$ will also admit $(A,-B)$ as a solution. This just corresponds to the complex conjugate solution of the wave equations. Hence, $A_0(\hat{p},\vec{x})=A_0(-\hat{p},\vec{x})$ and we get~\footnote{Terms that mix creation operators on opposite directions, that were estimed to be posible on general ground in \cite{nosotros}, are in fact absent.}
 \be
{\cal H}_{\textrm{loc}}=\int d^{d-1}p\, \frac{2\pi}{A_0(\hat{p},\vec{x})}  \, |\vec{p}|\, a^\dagger_{\vec{p}}a_{\vec{p}} \,.
\ee

 This shows the modular Hamiltonian acts as a thermal state with direction dependent temperature for local excitations. The inverse temperatures are given by
\be
\beta(\hat{p},\vec{x})= \frac{2\pi}{A_0(\hat{p},\vec{x})}\,.\label{betata}
\ee

\subsection{Dirac field}

In this subsection, we will apply for the Dirac field the same procedure we used for the scalar field in the previous one. In general terms, the eikonal approximation leads to the same geometric problem in the fermion case, hence we will frequently refer to the results above. However, we will emphasize the differences between the two cases due to the spinor nature of the solutions of (\ref{edir}). 

We start parameterizing the solutions $S(x)$ as in (\ref{sol_eik}), but in this case $g(x) \in \mathbb{C}^n$ is a spinor valued function ($n=2^{\lfloor \frac{d}{2} \rfloor}$). $S(x)$ satisfies the Euclidean Dirac equation, but each spinor component also satisfies the Euclidean Klein-Gordon equation as (\ref{eik_ekg}). Then, clearly, we have the same eikonal equations (\ref{eikonal_scalar_2}) for the function $\alpha=a+ib$. The first difference between the bosonic and fermionic case, is that the boundary conditions (\ref{eik_bc}) must be replaced by
\bea
&& a(0^+,\vec{x}) = a(0^-,\vec{x}) - 2 \pi s \hspace{0.5 cm} \forall \vec{x}\in V \,,\\
&& b(0^+,\vec{x}) = b(0^-,\vec{x}) - \pi . 
\eea
However, since we are interested in the limit of large $s$ (eikonal limit), the factor $(-1)$ in the spinors that leads to the jump $\pi$ in  $b$ in the second expression is a subleading effect as compared to the jump of order $s$ in $a$, and we can drop the factor $\pi$ on the $b$ function, recovering the same eikonal equations we obtained for the scalar field.

\subsubsection{Eigenfunctions and normalization}

As we claim above for the scalar, the solutions of the eikonal problem for the Dirac field analogous to (\ref{eik_sol}) are, 
\be
u^k_s(\vec{x}) =S^k_s(0^+,\vec{x}) = g_k(\vec{x}) \,  \mathrm{e}^{i s \int_{x_*}^{\vec{x}} \, \vec{B}_k(\vec{y}) \cdot d\vec{y}} \, . \label{u_dirac}
\ee
Normalizing these spinors as we did in the previous subsection for the scalar functions, we arrive at
\be
\langle u^k_s | u^{k^\prime}_{s^\prime}\rangle= \int_\Sigma d^{d-1}\sigma \, g_k^{\dagger} g_k \, J^{-1} \, \mathrm{e}^{-i \Delta s \, \sigma_{d-1} }\, \mathrm{e}^{-i\, s \,\sigma_i   \Delta k_i }=\delta(s-s')\delta(k-k') \, ,
\ee
which implies that $g^{\dagger}_k(\vec{x}) g_k(\vec{x}) = \frac{|s|^{d-2} \, J(\vec{x})}{(2 \pi)^{d-1}}$.

 This relation only determines the norm of the spinor $g$, and we need another formula to completely determine it. For that, we use that the solution $S(x)$ satisfies the Dirac equation,
\be
0= (\gamma_\mu \partial_\mu + m) S(x) = \mathrm{e}^{\alpha} [ (\gamma_\mu \partial_\mu g) + (\gamma_\mu \partial_\mu \alpha) g + m g] \implies (\gamma_\mu \partial_\mu \alpha) g = 0 \, .
\ee
The last implication is valid in the eikonal limit in which the derivatives of the exponent $\alpha$ are much greater than the ones of $g$ or the field mass. Equivalently,
\be
(\gamma_0  A_0 + i \, \vec{\gamma} \cdot \vec{B}) g = 0 \, , 
\ee
where $\vec{\gamma}=(\gamma_1, \ldots , \gamma_{d-1})$. Recalling the relation between the Euclidean gamma matrices $\{\gamma_\mu\}_{\mu=0,\ldots , d-1}$ and the Minkowskian gamma matrices $\{ \gamma^\mu_M \}_{\mu=0, \ldots, d-1}$, we can rewrite the last expression as
\be
\big(\gamma^0_M  A_0 +  \vec{\gamma}_M \cdot \vec{B} \big) g = 0 \, . \label{diti}
\ee

$A_0$ plays the role of the energy and $\vec{B}$ the one of the momentum in plane wave solutions.  Since $|A_0|=|\vec{B}|$, the spinor $g$ satisfies the same expression that the spinors of the plane wave solutions of the Dirac massless equation. 
Here we notice a difference with the scalar case. In the case of the scalar we take $s>0$, and in consequence $A_0>0$. However, for the Dirac field $s$ and $-s$ are independent time-reflected solutions, and each will be accompanied with $A_0$ of the same sign as $s$,\footnote{If $\vec{A}$, $\vec{B}$ is a solution with $s$, $-\vec{A}$, $\vec{B}$ is a solution with $-s$.} which corresponds to positive and negative energy solutions of the Dirac equation (\ref{diti}). 

A basis of spinors $\{w^j(\vec{p})\}_{j=1, \ldots, \frac{n}{2}} \subset \mathbb{C}^n$ which satisfy the Dirac equation with positive energy
\be
\big( \gamma^0_M  |\vec{p}| -  \vec{\gamma}_M \cdot \vec{p} \big) w^j(\vec{p}) = 0 \, , 
\ee
 can be chosen such $w^{j\dagger}(\vec{p})w^{j^{\prime}}(\vec{p})=2 |\vec{p}| \, \delta_{j,j^{\prime}}$. Then the desired solutions for the spinor $g$ are
\be
g^j_k(\vec{x}) = \sqrt{ \frac{|s|^{d-2} \, J(\vec{x})}{(2 \pi)^{d-1} \, 2 | A^k_0(\vec{x})| }} \, w^j\big( - \mathrm{sgn}(s) \, \vec{B}({\vec{x}})\big) \, , \label{b_dirac}
\ee
where we have used the fact that $\mathrm{sgn(s)}=\mathrm{sgn(A_0)}$. Therefore, on top of the $k$ degeneracy, there is a $n/2$ spinor degeneracy for each $s$, which can have two signs. 
 
\subsubsection{Local terms in the modular Hamiltonian}
Now, we are ready to explicitly calculate the Hamiltonian kernel $\mathcal{H}(\vec{x}' , \vec{y}')$ replacing (\ref{u_dirac}) and (\ref{b_dirac}) in (\ref{H_ker}). Doing the same near point approximation ($\vec{x}' = \vec{y}' + \Delta \vec{x}'$) and the same change of variables $k \leftrightarrow \vec{B}_{\vec{x}}$ as we did for the scalar field, we get
\be
H_{\textrm{loc}}(\vec{x}',\vec{y}') \simeq  \int_{\mathbb{R}} ds\,\int_{\cal K} d^{d-2}k \,\, (-2 \pi \, s)  \,  \frac{|s|^{d-2} \, J(\vec{x})}{(2 \pi)^{d-1} \, 2 | A^k_0(\vec{x})| } \, \sum^{\frac{n}{2}}_{j=1}w^j\big( - \mathrm{sgn}(s) \, \vec{B}({\vec{x}})\big) w^{j\dagger}\big( - \mathrm{sgn}(s) \, \vec{B}({\vec{x}})\big)\, \mathrm{e}^{i \,s\,\vec{B}_k(\vec{x}) \cdot \Delta\vec{x}'}  \, . \label{Kernel_H}
\ee

Using the identity
\be
\sum^{\frac{n}{2}}_{j=1}w^j(\vec{p}) w^{j\dagger}\big(\vec{p})= |\vec{p}| +\gamma_M^0 \vec{\gamma_M} \cdot \vec{p}
\ee
we can write equation (\ref{Kernel_H}) as
\be
H_{\textrm{loc}}(\vec{x}',\vec{y}') \simeq  \int_{\mathbb{R}} ds\,\int_{\cal K} d^{d-2}k \,\, (-2 \pi \, s)  \,  \frac{|s|^{d-2} \, J(\vec{x})}{(2 \pi)^{d-1} \, 2 | A^k_0(\vec{x})| } \, \Big( | A^k_0(\vec{x})| - \mathrm{sgn}(s) \, \gamma^0_M \vec{\gamma}_M  \cdot \vec{B}^k(\vec{x}) \Big) \, \mathrm{e}^{i \,s\,\vec{B}_k(\vec{x}) \cdot \Delta\vec{x}'}  \, . 
\ee

Next we apply the same change of integration variables as in (\ref{changee}) where we have to take $s$ positive, $s\rightarrow |s|$, keeping in mind we have two solutions with opposite signs of $s$ for the same $\vec{p}$
\bea
H_{\textrm{loc}}(\vec{x}',\vec{y}') &\simeq& \sum_{\mathrm{sgn}(s)} \int \frac{d^{d-1}p}{(2\pi)^{d-1}}  \,\, \frac{-2 \pi \, \mathrm{sgn}(s)\,|s(\vec{p})|}{2 | A_0(\hat{p},\vec{x})| } \, \Big( | A_0(\hat{p},\vec{x})| - \frac{\mathrm{sgn}(s)}{|s|} \, \gamma^0_M \vec{\gamma}_M  \cdot \vec{p}_{\vec{x}} \Big) \, \mathrm{e}^{i \,\vec{p}_{\vec{x}} \cdot \Delta\vec{x}'}  \nn \\
&=& \int \frac{d^{d-1}p}{(2\pi)^{d-1}} \,\, \frac{2 \pi}{|A_0(\hat{p},\vec{x})|}   \, (\gamma^0_M\vec{\gamma}_M\cdot \vec{p})\, \mathrm{e}^{i \,\vec{p} \cdot \Delta\vec{x}'}  \, . \label{Kernel_H2}
\eea
Hence, the sum over the two values of $s$ amounts to summing over positive and negative energy solutions of the Dirac Hamiltonian, which we recognize within the brackets in the last expression. 

The operator expression for the local modular Hamiltonian follows by smearing the kernel $H(\vec{x}',\vec{y}')$ with the field operators as in (\ref{modH_dirac}). Using the explicit expression of the fields in terms of the creation $\{a^{\dagger}_j(\vec{p}) , b^{\dagger}_j(\vec{p}) \}$ and annihilation operators $\{ a_j(\vec{p}) , b_j(\vec{p}) \}$ in Fock space
\be
\psi(\vec{x}) = \int_{\mathbb{R}^{d-1}} \frac{d^{d-1}p}{\sqrt{(2\pi)^{d-1}\,2|\vec{p}|}} \Bigg( \sum_{j=1}^{\frac{n}{2}} u^j( \vec{p} ) \, a_j(\vec{p}) \, \mathrm{e}^{i \vec{p} \cdot \vec{x}} + v^j( \vec{p} )\, b^{\dagger}_j(\vec{p}) \,\mathrm{e}^{- i \vec{p} \cdot \vec{x}} \Bigg) \, . 
\ee
This amounts to replace the Dirac Hamiltonian within the brackets in (\ref{Kernel_H2}) by its second-quantized version, 
\be
{\cal H}_{\textrm{loc}} = \int d^{d-1}p \, \frac{2 \pi |\vec{p}|}{|A_0(\hat{p},x)|}  \sum_{j=1}^{\frac{n}{2}}  \Big( a^{\dagger}_j(\vec{p}) a_j(\vec{p}) + b^{\dagger}_j(\vec{p}) b_j(\vec{p}) \Big) \, .
\ee
Here we see exactly the same result as for a scalar. The local temperature is direction dependent and given by the same formula (\ref{betata}).\footnote{In order to obtain the local temperatures is only necessary to consider the eikonal solution with $s>0$, $A_0>0$, as in the scalar case.}  

\section{Summary: Universality of local temperatures}
Summarizing the results, we have that for an arbitrary region $V$ the local inverse temperatures can be obtained by solving the purely geometric problem given by the eikonal equations (\ref{5}-\ref{6}), (\ref{7}-\ref{8}), for two vector fields $A$ and $B$ on Euclidean space. These equations imply the two vector fields are orthogonal, have equal modulus, and both of them are gradients except for $A$ on $\partial V$, where it has a magnetic flux-type source along the boundary. This singularity means the circulation of $A$ around $\partial V$ in the positive time direction when crossing $V$ is $2\pi$. Further, $A$ is orthogonal to $V$ and $B$ is parallel to $V$. The same eikonal solutions in Euclidean space also give the local temperatures for the complementary region $\bar{V}$.    

On $x_0=0$, $\vec{x}\in V$, the vector $A(x)$ is purely time-like and future directed, $\vec{A}(x)=0$, $A_0(x)>0$, and the field $B(x)$ is space-like, $B_0(x)=0$. Given a point $\vec{x}\in V$ at $x_0=0$, and a particular direction $\hat{p}$, we have to find the eikonal solution with $B(\vec{x})$ pointing in this same direction, $\hat{B}(\vec{x})=\hat{p}$, and the value of $A_0(\vec{x})=|A(\vec{x})|=|B(\vec{x})|$ gives us the temperature as 
\be
T(\hat{B}(\vec{x}),\vec{x})=\frac{A_0(\hat{B}(\vec{x}),\vec{x})}{2\pi}\,,
\ee
or the inverse temperature
\be
\beta(\hat{B}(\vec{x}),\vec{x})=\frac{2\pi}{A_0(\hat{B}(\vec{x}),\vec{x})}\,.
\ee

This result is the same for free scalars and fermions, independently of the mass, pointing to some universality of the local temperatures across different theories. As by its very definition the local temperatures are defined for a regime of large energies we expect the same result would apply to super-normalizable theories as well.   

In general the local temperatures are direction dependent; we will see an explicit example in section \ref{slab}. A general result is that the temperatures in a direction and the opposite one are the same,
\be
\beta(\hat{B}(\vec{x}),\vec{x})=\beta(-\hat{B}(\vec{x}),\vec{x})\,. 
\ee
This is a technical consequence of the fact that if $(A,B)$ is a solution, $(A,-B)$ is also a solution, and can be traced to time inversion invariance of both the vacuum state and the region $V$. These imply that the relative entropy between the excitation with momentum $\vec{p}$ and the vacuum do not change if we invert the momentum. This will not hold for points in the causal development of $V$ outside $x_0=0$.   

Unfortunately, we do not know of a general method to solve these equations. In the next sections we show some particular analytic solutions. However, these differential and algebraic equations are in principle solvable numerically. In contrast to ordinary eikonal equations for Schroedinger or Maxwell equations, which give place to particle-like trajectories with local action, and hence boil down to ordinary differential equations, in the present case the eikonal equations are much more non local. The solutions cannot be found locally without taking into account boundary conditions imposed in $\partial V$ far away. 
   
These same technical difficulties have impeded us from proving some natural expectations. For example,
 for any region we expect $\beta(\hat{p},\vec{x})>0$ to exist and be different from zero for any direction $\hat{p}$ and point $\vec{x}\in V$ outside $\partial V$. This points to the existence of sufficiently many solutions of the eikonal equation such that we can fix the direction of the field $B$ at a point, $\hat{p}=\hat{B}(\vec{x})$. Further, we expect this solution with $\hat{B}(\vec{x})$ fixed to be unique, but this is less obviously necessary. If there would be more than one solution we would have to sum over the different $2\pi/A_0(\hat{B}(\vec{x}),\vec{x})$ in the modular Hamiltonian to obtain the inverse temperature $\beta$. Another necessity is that $\beta(\hat{B}(\vec{x}),\vec{x})$ for fixed $\vec{x}$ and $\hat{B}(\vec{x})$, as a function of the region, has to be increasing   
 under inclusion of regions \cite{nosotros}. This is a consequence of monotonicity of relative entropy.    
  
\subsection{Conformal transformations}
The eikonal equations come from high energy solutions and are independent of mass. Then, they must be covariant under conformal transformations. This allows us to find the solutions in a conformally transformed space from solutions in the original space. To see the rules of transformations is convenient to note that a solution of the massless Klein-Gordon equation will transform with a pre-factor, but this factor will not change the term linear in $s$ in the exponent. Therefore the exponent $\alpha=a+i b$ only undergoes a coordinate transformation. 

To see this more explicitly, let $\tilde{x}=\tilde{x}(x)$ be a point transformation between manifolds ${\cal M}$ and $\tilde{{\cal M}}$, with 
\be
\tilde{g}_{\mu\nu}=\Omega^2 g'_{\mu\nu}\,,\hspace{1cm}g'_{\mu\nu}=\frac{\partial x^\alpha}{\partial \tilde{x}^{ \mu}}\frac{\partial x^\beta}{\partial \tilde{x}^{ \nu}} g_{\alpha\beta}\,.\label{conf}
\ee
$A$ and $B$ will transform as vectors,
\bea
\tilde{A}_\mu &=& \frac{\partial x^\alpha}{\partial \tilde{x}^{ \mu}} A_\alpha \,,\\
\tilde{B}_\mu &=&  \frac{\partial x^\alpha}{\partial \tilde{x}^{ \mu}} B_\alpha \,.
\eea
It is immediate to check using (\ref{conf}) that the relations $\tilde{A}\cdot \tilde{B}=0$, $|\tilde{A}|=|\tilde{B}|$, hold in the new space. They are also gradients on the new coordinates and the circulation of $\tilde{A}$ around $\partial \tilde{V}$ does not change because $a$ is preserved on the two sides of the cut. 
   
\section{Simple solutions}
In this section we show some examples of solutions of the eikonal equations that are particularly simple to obtain. In both 
of these examples however, the local temperatures are direction independent, and given by a contribution proportional to the stress tensor in the modular Hamiltonian. 
\subsection{Two dimensions}
In $d=2$ we have outside the boundary of the region the equations 
\bea
(\partial_1 a)^2+(\partial_2 a)^2 &=&(\partial_1 b)^2+(\partial_2 b)^2\,,\\
\partial_1 a \, \partial_1 b+\partial_2 a \, \partial_2 b &=&0\,.
\eea
These are equivalent to
\bea
\partial_1 a=\pm \partial_2 b\,,\\
\partial_2 a=\mp \partial_1 b\,.
\eea
These are exactly the Cauchy Riemann equations for the function $\alpha=a+i b$, that implies it is either analytic or anti-analytic.\footnote{In geometric terms, every analytic function gives place
 to an orthogonal coordinate system where the gradients of the coordinates have equal modulus, $|A|=|B|$. Note however that it is possible that $|A|=|B|=0$ for some points. 
This is  shown in \cite{cardy} for the case of two intervals. This is not a problem for the eikonal equations because these special points do not appear in the region $V$.}
Analyticity is a great help which allows us to solve the problem in full generality. For 
a region consisting in $n$ disjoint intervals $V=\cup_{i=1}^n (l_i,r_i)$, we need to impose the 
right boundary conditions (\ref{eik_bb}-\ref{eik_bc}) on the cut $V$, and to impose $\alpha$ to be analytic or anti-analytic outside $V$. Writing $z=x_1+i x_0$, this is solved by the functions (already introduced in (\ref{funi})),
\bea
\alpha_1(z)=i s \log\left(\prod_{i=1}^n\frac{z-l_i}{z-r_i}\right)\,, \label{funi1}\\
\alpha_2(z)=-i s \log\left(\prod_{i=1}^n\frac{\bar{z}-l_i}{\bar{z}-r_i}\right)\,.
\eea  
Multiplying these functions by an analytic (or anti-analytic) function we obtain another solution of the eikonal equations. However, these functions would contain other singularities or be unbounded at infinity. 

\begin{figure}[t]
\begin{center}
\includegraphics[width=0.45\textwidth]{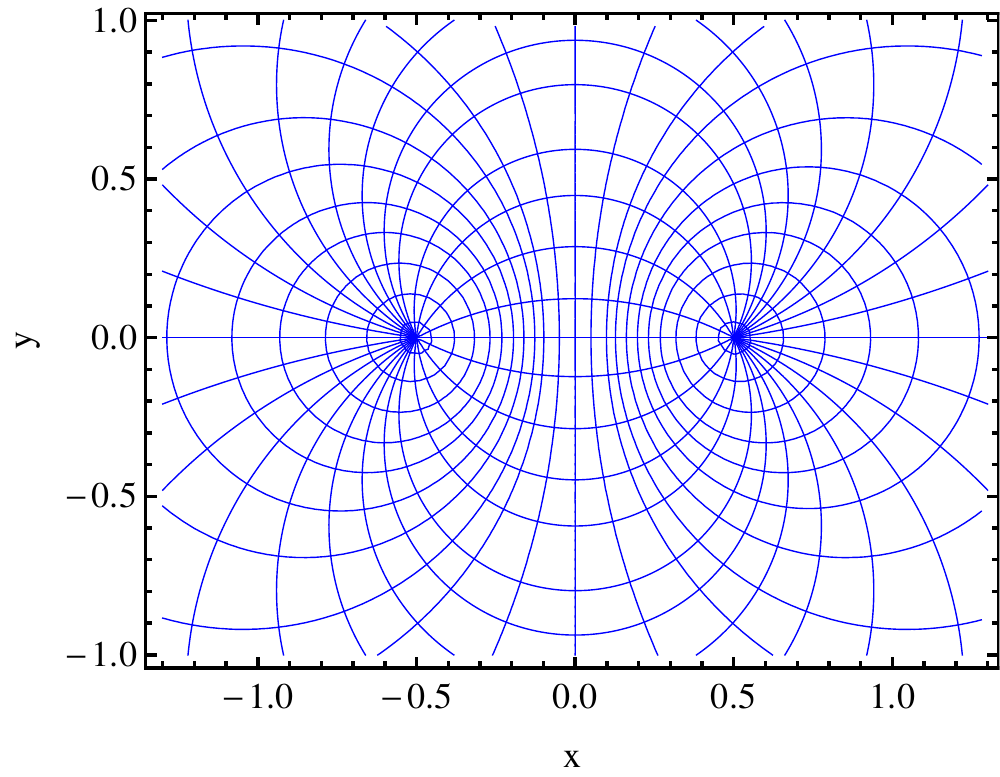} 
\caption{The lines of constant $a$ and $b$ of the solution (\ref{funi1}) of the eikonal problem corresponding to an interval $(-1/2,1/2)$ in $d=2$. The lines form an orthogonal coordinate system. Curves of constant $a$ are circles passing through the end-points of the interval, while curves of constant $b$ are circles around these end-points.}
\label{eiko2}
\end{center}
\end{figure}

Hence, $\alpha_1, \alpha_2$, which are the conjugate of one another, form the full space of solutions for $\alpha$. An example for a single interval is shown in figure \ref{eiko2}. On $V$ the vector $B$ has non-zero component
\be
B_1(x_1)=\partial_1 \Im (\alpha/s)= \pm \sum_i\left(\frac{1}{x_1-l_i}+\frac{1}{r_i-x_1}   \right)\,,
\ee
where the plus (minus) sign corresponds to the solution $\alpha_1$ ($\alpha_2$). Then, in order to compute the local temperature at a point $x_1\in V$ we have to choose one of the two possible directions for $B_1$. This amounts to choose either of the solutions. In the present case both directions give place to the same temperatures since $B_1$ differ in sign but not in modulus. We have
\be
\beta(x)=\frac{2 \pi}{A_0} =\frac{2 \pi}{|B_1|}=2 \pi\,\left(\sum_i\left(\frac{1}{x_1-l_i}+\frac{1}{r_i-x_1}   \right)\right)^{-1} \,.
\ee 
This is the result quoted in (\ref{timeo}). It was first obtained in \cite{diracd2} through an exact calculation of the modular Hamiltonian for massless free Dirac fields in $d=2$, but regarding the local terms, it also holds for massive free fermions and free massive scalars \cite{nosotros}. As the two temperatures in both directions have to coincide, the local term in ${\cal H}$ is proportional to the energy density operator $T_{00}$.  

\subsection{Rindler space}
\label{dfgh}
In Rindler space (corresponding to $V=\{\vec{x}/ x_1>0\}$) we have that the exact modular Hamiltonian for any theory is
\be
{\cal H}=2 \pi \int_V d^{d-1}x\, x_1 T_{00}(\vec{x})\,. 
\ee
Hence $\beta(\vec{x})= 2 \pi x_1$, independently of direction. However, it is instructive to obtain this result using the eikonal equations.

We take cylindrical coordinates $r, \theta, \vec{x}_\parallel=\{x_3,\cdots,x_{d-1}\}$ with $x_1=r \cos(\theta)$, $x_0=r \sin(\theta)$. Due to the rotational and translational symmetries of the problem in Euclidean space we take   
\bea
A&=& \frac{\hat{\theta}}{r}\,,\\
B&=&f(r) \hat{r} + \vec{k}   \,, 
\eea
with $\vec{k}=\{0,0,k_1,\cdots,k_{d-1}\}$, and $\hat{\theta}, \hat{r}$, the unit vectors proportional to the gradient of the coordinates. This corresponds to a wave solution obtained by separation of variables of the form
\be
S\sim e^{a+i b}=e^{s \,\theta +i \,s \,\vec{k}\cdot \vec{x}_\parallel+ i \,s\, h(r)}\,, 
\ee
with $h'(r)=f(r)$.
It follows from $|A|=|B|$ that
\be
f(r)=\pm\sqrt{\frac{1}{r^2}-\vec{k}^2}\,.\label{114}
\ee
If $r>|\vec{k}|^{-1}$ we have an imaginary $f(r)$ and $B$. In that case we could pass the part $f(r) \hat{r}$ of $B$ to $A$, but then $A$ would not be normal to $V$. We see that the solution has a cylindrical domain given by $r<|\vec{k}|^{-1}$ and cannot be extended to all space.  
 Physically, this is because for $r>|\vec{k}|^{-1}$, in the eikonal limit $s\rightarrow \infty$, the wave solution is exponentially damped with a large $s$ factor in the exponent, and has to be considered zero in this range. This is no impediment to get a solution for any $\vec{x} \in V$ with arbitrary direction $\hat{B}(\vec{x})$. 
 
Taking a point $\vec{x}$ with $\theta=0$, $x_1=r$ on $V$ we have 
\be
B(\vec{x}) = f(x_1) \hat{x}_1 + \vec{k}\,.
\ee
Note that for fixed $x_1$ we can take solutions with the two signs in (\ref{114}), and with different $\vec{k}$, having $|\vec{k}|$ ranging from $0$ to $x_1^{-1}$. In this way we can choose any direction for $\hat{B}(\vec{x})$. There is exactly one solution for each direction $\hat{B}(\vec{x})$ at a point $\vec{x}\in V$.\footnote{Though this set of solutions for a fixed $\vec{x}$ do not exhaust the full set of solutions since there are solutions with range not including the point $\vec{x}$.}   
However, in this particular problem for the Rindler wedge, for any choice of solution we have
\be
A_0 = |B|=x_1^{-1}\,,
\ee  
independently of the direction of $\hat{B}(\vec{x})$.
Therefore, as expected, 
\be
\beta(\hat{B}(\vec{x}),\vec{x})=\frac{2\pi}{A_0(\hat{B}(\vec{x}),\vec{x})}=2 \pi x_1\,.
\ee

\section{Local temperatures for a wall}
\label{slab}

In this section we compute explicitly the local temperatures for a region with the form of a wall in $d$ dimensions. Without loss of generality we set the width $L$ of the wall  $V$ to $L=1$ in the $x_1$ direction and unlimited in the other spatial directions, i.e., we take  $V=\{\{0,x_1,\cdots,x_{d-1}\}\,/\, x_1\in (-1/2,1/2)\,,\,\,x_i\in (-\infty,\infty)\,,\,\,i=2,\cdots,x_{d-1}\}$.  

We will not solve the eikonal equations directly, but instead will first derive further identities for the solutions $S$ of the Klein-Gordon equation that apply to the present case due to the particular symmetries of the problem. We then make the eikonal approximation. The results for the eikonal variables and the local temperatures will be given in terms of solutions of algebraic equations of fourth degree. 

By separation of variables we can dimensionally reduce the problem to a massive one  in $d=2$.
For ease of notation  we will call $x=x_1$, $y=x_0$, $\vec{x}_\parallel=\{x_3,\cdots,x_{d-1}\}$.
 We take for the solution of the wave equation
\be
S(x,y) \, e^{i \,s\, \vec{k}\cdot \vec{x}_\parallel }\,,
\ee
with $\vec{k}=\{0,0,k_1,\cdots,k_{d-1}\}$. 
 The equation for the function $S$ in $d=2$ is
\be
(-\nabla^2+ s^2 k^2) \,S(x,y)=0\,,\label{masssa}
\ee
with $k=|\vec{k}|$. $S$ satisfies the boundary conditions (\ref{bcs}) for the two dimensional problem with $V$ given by the interval $I=(-1/2,1/2)$ of width $L=1$. Eq. (\ref{masssa}) corresponds to a massive field with $m=s \, k$, but in contrast to the discussion in section 2, here we are not allow to discard the mass invoking that we are interested in the high energy limit. This is because the momentum in the parallel direction to the wall can be as large as we want, and in particular it can be of the same order as $s$. To keep track that we are interested in high momentum as well as high $s$ we have defined the two dimensional mass to be $s \,k$, where $k$ will be ${\cal O}(1)$ in the eikonal parameter.

Now we make use of the techniques initially developed in \cite{myers} to solve the problem of scattering of waves from a metallic strip. These were adapted in \cite{scalar} to a slightly different version of the present problem where the multiplicative boundary condition on the interval is given by a phase instead of a real factor (that is, a problem with imaginary $s$). This problem is related to the Renyi entropies of a massive field in an interval. We derive the corresponding equations for the function $S$ in the Appendix \ref{ap2}. 

For a scalar field of mass $m$ in $d=2$ satisfying the Klein-Gordon equation with the multiplicative boundary conditions on an interval $(L_1,L_2)$, $L=L_2-L_1$, we have, writing $z=x+i y$,    
\be
\left((L_1-z)\,\partial_z-(L_2-\bar{z}) \,\partial_{\bar{z}}-\frac{t\, d'}{2\, d}\right)\, S= C_{00}\, \left(-\frac{i\, L}{d} \,\partial_z + \frac{2\, s \,d-i\, t\, d'}{2 \,(1+d^2)}\right) \bar{S}\,,  
\label{efi1}\ee 
supplemented by its conjugate equation
  \be
\left((L_1-\bar{z})\,\partial_{\bar{z}}-(L_2-z) \,\partial_{z}-\frac{t\, d'}{2\, d}\right)\, \bar{S}= C_{00}^*\, \left(\frac{i\, L}{d} \,\partial_{\bar{z}} + \frac{2\, s \,d+i\, t\, d'}{2 \,(1+d^2)}\right) S\,.\label{efi2}  
\ee 
The function $C_{00}(t)$ satisfies $|C_{00}|^2=d(t)^2+1$, and the real function $d(t)$ satisfies a non-linear ordinary differential equation of the Painlev\'e V type,  
\be
d''+\frac{d'}{t}-\frac{(1+2\, d^2)(d')^2}{d\,(1+d^2)}+\frac{(t^2 (1+ 2\, d^2) + (t^2 - 4\, s^2)\, d^4)}{
 t^2\, d\, (1 + d^2)} =0\,, \label{painleve0}
\ee
 with boundary conditions
\bea
d(t) &\sim &  -\frac{1}{2 \,s\,\left(\log(t)-\log(2)+2 \gamma_E+(\psi(i s)+\psi(-i s))/2\right)}   \,, \hspace{1cm} t\rightarrow 0\,, \label{boundary1}\\ 
d(t) & \sim &   \frac{\pi}{2\sinh (\pi \, s) K_{2 i s}(t)}  \,, \hspace{1cm} t\rightarrow \infty\,,\label{boundary2}
\eea
where $\gamma_E$ is the Euler constant, $K$ is the modifyed Bessel function, and $\psi$ the digamma function.\footnote{The function $d$ is related to the one called $u$ in \cite{scalar} by the change of variables $d=i\frac{t (1+u^2)}{u (1+2 i s)+t u' }$, or $u= \frac{2 s d -i t d'}{t (1+d^2)}$ , and taking $a\rightarrow -i\,s$ in that work.} 

Here we have to choose $L_1=-1/2$, $L_2=1/2$, $L=1$, and hence $t=m=s\, k$. 

\subsection{The eikonal limit for the Painlev\'e equation}
We first take the eikonal limit $s\rightarrow \infty$ of the solution of the Painlev\'e equation (\ref{painleve0}) with boundary conditions (\ref{boundary1}-\ref{boundary2}) that determine the coefficients of the equations for $S$ we are going to use. We take the limit of $s\rightarrow \infty$, $t=m L=m=s k\rightarrow \infty $, keeping $k$ fixed. Then, this is not directly related to the asymptotic limit $t\rightarrow \infty$ and $s$ fixed, that describes the boundary condition of the differential equation. The inspection of the solutions in this limit shows there are two regimes.

First, for $k<2$, the function $d$, which depends on $s$ and $t=s k$, converges to a finite value $f(k):=\lim_{s\rightarrow \infty} d_s(s k)$,\footnote{We are writing explicitly the $s$ dependence of $d(t)$ as $d_s(t)$.}  and in consequence its derivatives, $\lim_{s\rightarrow \infty}\frac{d}{dt} d_s(s k)=s^{-1} f'(k)$, $\lim_{s\rightarrow \infty}\frac{d^2}{dt^2} d_s(s k)=s^{-2} f''(k)$, go to zero with inverse powers of $s$. Plugging this information into the differential equation (\ref{painleve0}) gives an algebraic explicit solution for $d_s(t)$ 
\be
d_s(s k)\sim f(k)=\sqrt{\frac{k}{2-k}}\,, \hspace{.7cm} k<2\,, \,\,s\rightarrow \infty \,.\label{para1}
\ee

For $k>2$ instead, the asymptotic solution (\ref{boundary2}) for large $t$ is still valid for large $s$ and $t=k s$. We have using the asymptotics of the Bessel function \cite{asymptotic} an exponentially increasing $d_s(s k)$,  
\be   
d_s(s k)\sim  \textrm{cons} \,\sqrt{s}\, e^{\sqrt{k^2-4}\, s } \, \,, \hspace{.7cm} k>2\,, \,\,s\rightarrow \infty \, . \label{para2}
\ee
It can be checked that this expression (as the one in (\ref{para1}) for $k<2$) also solves the Painlev\'e equation at leading order in $s$ for $k>2$. 
With this behaviour all derivatives are of the same order, and we have $d'(t)/d(t)\sim \frac{\sqrt{k^2-4}}{k}$.

\subsection{Eikonal limit for the equations of $S(z)$ and angle dependent temperatures}

\begin{figure}[t]
\begin{center}
\includegraphics[width=0.5\textwidth]{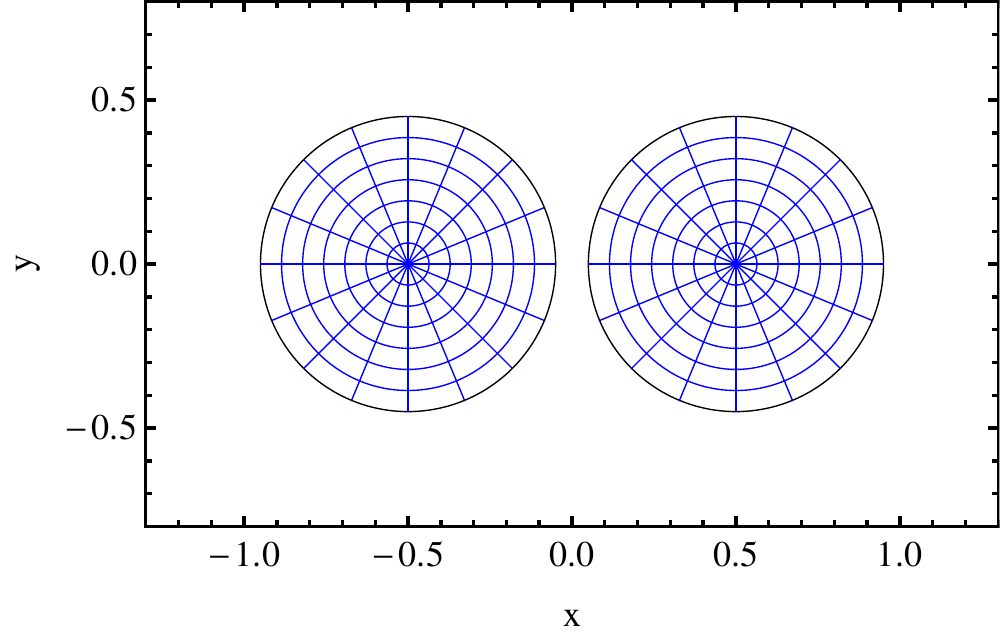} 
\caption{The lines of constant $a$ and $b$ corresponding to the solution of the eikonal problem  for a wall in the case $k>2$ (in this figure $k=20/9$). The solution corresponds to two non-overlapping Rindler solutions, one around each of the interval end-points.  }
\label{eiko3}
\end{center}
\end{figure}

Now we apply the eikonal limit to the equations for $S$ and $\bar{S}$ (\ref{efi1}-\ref{efi2}). Both $S$ and $\bar{S}$ will have a component proportional $e^{\alpha}$ and another one proportional to  $e^{\alpha^*}$. In the large $s$ limit these two components have to satisfy the equations (\ref{efi1}-\ref{efi2}) independently because the different large phases cannot be coherent in any small region. Hence, we can replace $S$ by $g_1 e^{\alpha}$ and $\bar{S}$ by $g_2 e^{\alpha}$, where $g_1,g_2$ are slowly varying functions of the position. We then set 
\bea
\partial_x S &=& s\, X \,S\,,\hspace{.7cm}  \partial_x \bar{S}=s\, X\, \bar{S}\,,\\
\partial_y S &=& s \,Y\, S\,,\hspace{.7cm}  \partial_y \bar{S}=s\, Y\, \bar{S}\,,
\eea  
where
\be
X=A_x + i\, B_x\,,\hspace{.7cm} Y=A_y + i \, B_y\,,\label{vectorial}
\ee
are two complex numbers with the information of the $x,y$ components of $\vec{A}$ and $\vec{B}$. We also have $A_\parallel=0$, $B_\parallel=\vec{k}$. 
Replacing this into (\ref{efi1}-\ref{efi2}), and expanding to the leading order in $s$, and get two complex algebraic equations linear in $S$ and $\bar{S}$. The determinant of this system of linear equations must vanish and this gives a complex equation for $X$ and $Y$. This is supplemented by the Klein-Gordon equation
\be
X^2+Y^2=k^2\,, \label{suple} 
\ee
to give the complete solution for the vectors $\vec{A}$ and $\vec{B}$.

\begin{figure}[t]
\begin{center}
\includegraphics[width=0.55\textwidth]{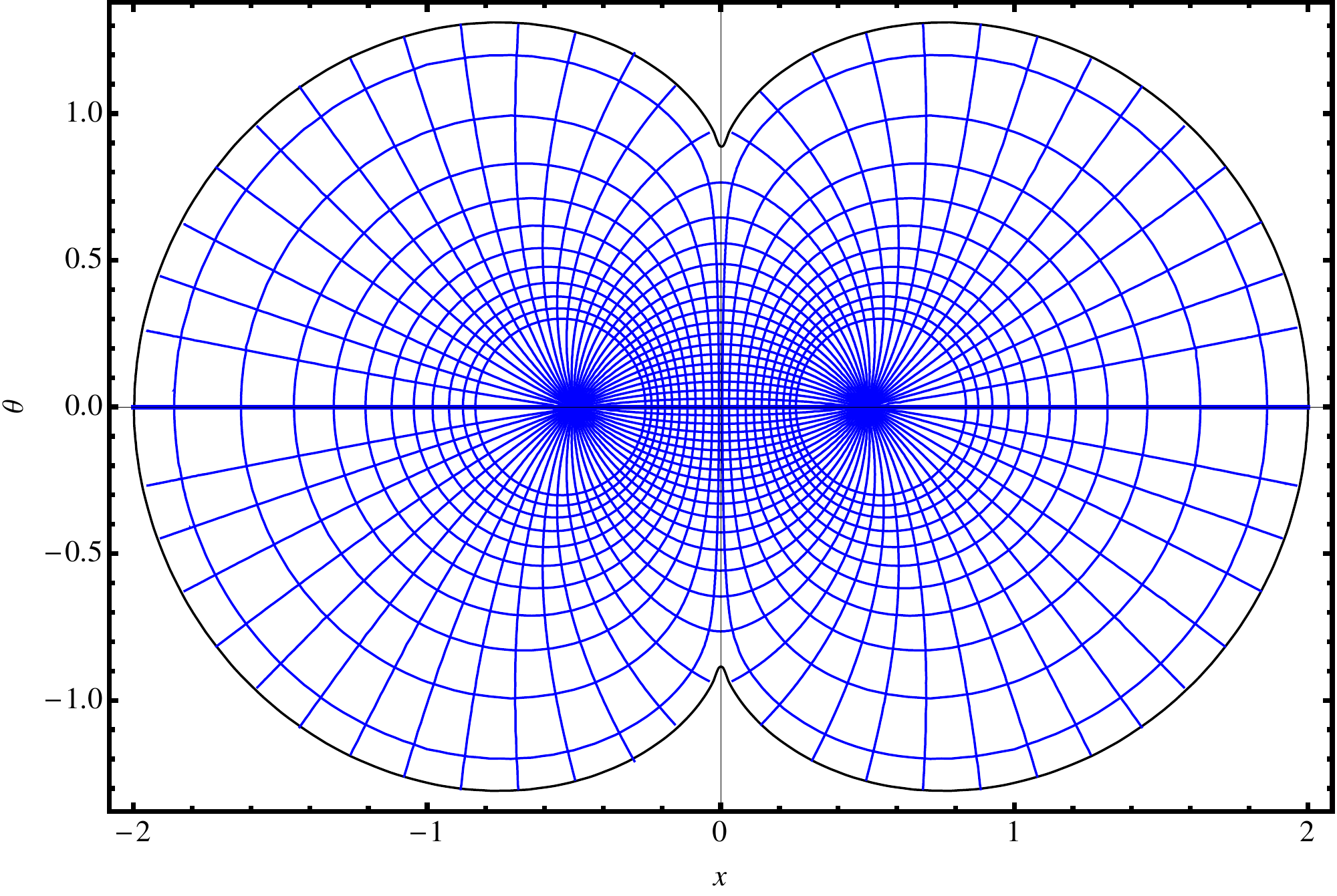} 
\caption{The orthogonal lines of constant $a$ and $b$ for the solution of the eikonal problem corresponding to a wall in the case $k=1/2$. The domain of the solution is a bounded region in the plane (bounded by the black curve), but it is not disjoint, and in this sense is intermediate between the disjoint bounded solution for $k>2$ (figure \ref{eiko3}) and the massless solution $k=0$ that occupy all the plane (figure \ref{eiko2}).}
\label{eiko}
\end{center}
\end{figure}

As explained above, there are two regimes for the large $s$ expansion of the coefficients of the equations.

\subsubsection*{k>2}
 For $k>2$, taking into account (\ref{para2}), we get
\be
  Y+ y^2\, X^2+ (x^2-1/4)\, Y^2  - 2 x y \,X Y  =1\,.\label{dyp}
\ee
This, supplemented with (\ref{suple}), give us four solutions for each point $(x,y)$. Two of them are purely real and have to be discarded. Another duplication corresponds to complex conjugated solutions, $\vec{B}\rightarrow -\vec{B}$. There are complex solutions only for $|\vec{x}-L_1|<k^{-1}$ or $|\vec{x}-L_2|<k^{-1}$, that is, inside two circles of radius $k^{-1}$ around the two end-points of the interval. This is analogous to the case of the Rindler space. In fact, the solutions inside these circles exactly coincide with Rindler solutions. The contours for $a,b$ are illustrated in the figure \ref{eiko3}.

This situation can be easily understood. The Rindler solutions around each interval end-point has domain restricted to a radius $k^{-1}$ (see section \ref{dfgh}). For $k>2$, which represents a high enough mass or parallel momentum, these Rindler solutions do not overlap, the respective domains do not reach to the mid-point of the interval at the origin. Hence, these are genuine solutions of the eikonal problem on the wall. This will change for $k<2$ where the Rindler solutions have a non zero overlap.

\begin{figure}[t]
\begin{center}
\includegraphics[width=0.55\textwidth]{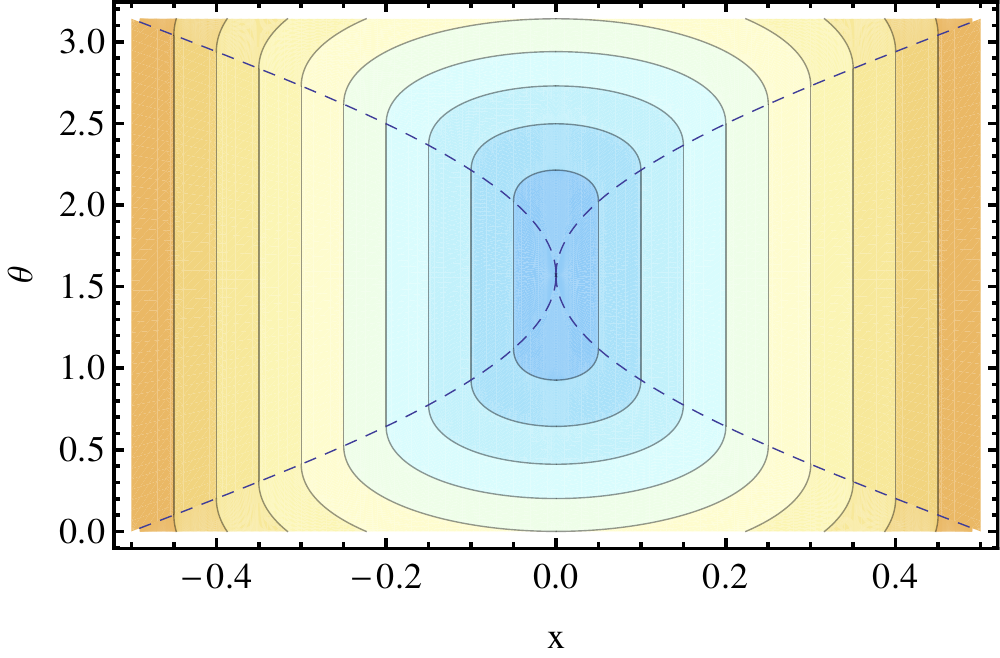} 
\caption{Contour plot of $\beta(x,\theta)$. The contour curves are equally spaced in $\beta$. Warmer colors indicate higher temperatures. The dashed lines separate the two regimes $k>2$ (the two regions attached to the end-points $x=\pm 1/2$), and $k<2$. For $k>2$, $\beta$ is independent of $\theta$ and coincides with Rindler result corresponding to the plane nearest to $x$. The $k<0$ solution which describes $\beta$ for the rest of the $x,\theta$ plane does show angle dependence of temperatures for fix $x$.}
\label{temp}
\end{center}
\end{figure}

Computing the solutions on $y=0$ on the interval gives for the non zero components
\bea
B_x &=&\pm\frac{\sqrt{1-k^2 (x+1/2)^2}}{ x+1/2}\,,\hspace{.7cm} A_y = (1/2 +  x)^{-1}\,,\hspace{.7cm} -1/2<x<-1/2+k^{-1}\,.\\
B_x &=&\pm\frac{\sqrt{1-k^2 (1/2-x)^2}}{ 1/2-x}\,,\hspace{.7cm} A_y = (1/2 -  x)^{-1}\,,\hspace{.7cm} 1/2-k^{-1}<x<1/2\,.
\eea
This gives Rindler-like inverse temperatures $2\pi/A_y$ independent of $k$, and hence independent of angle in the plane of the wall, but in contrast to the Rindler case, this situation is valid for a restricted range of angles. Let us compute the angle $\theta\in(0,\pi)$ from the positive direction of the $x$ axes,
\be
\cos(\theta)= \frac{B_x}{\sqrt{B_x^2+k^2}}=\frac{B_x}{A_y}\,,\hspace{.7cm} \sin(\theta)=\frac{k}{A_y}\,.
\ee
The solution with $k>2$, having Rindler-like temperatures, is restricted to the range
 \bea
\beta(x,\theta) &=&2 \pi \,(1/2 +  x)\,,\hspace{.2cm} -1/2<x<0\,, \hspace{.3cm} \acos(2 \sqrt{-x (1 + x)}) <\theta<  \acos(-2 \sqrt{-x (1 + x)}) \,.\\
\beta(x,\theta) &=&2 \pi\, (1/2 -  x)\,,\hspace{.4cm} 0<x<1/2\,, \hspace{.4cm}\acos(2 \sqrt{x (1 - x)}) <\theta<  \acos(-2 \sqrt{x (1 - x)})  \,.
\eea
That is, this solution holds for directions that are enough away from the $x$ axes. A contour plot of $\beta(x,\theta)$ showing the range of the solution is shown in figure \ref{temp}. Moreover in figure \ref{tempout} we show the temperature in the region outside the strip, in particular we plot the $x>1/2$ zone but the same behaviour is obtained in the reflected  $x<-1/2$ region.

\begin{figure}[t]
\begin{center}
\includegraphics[width=0.5\textwidth]{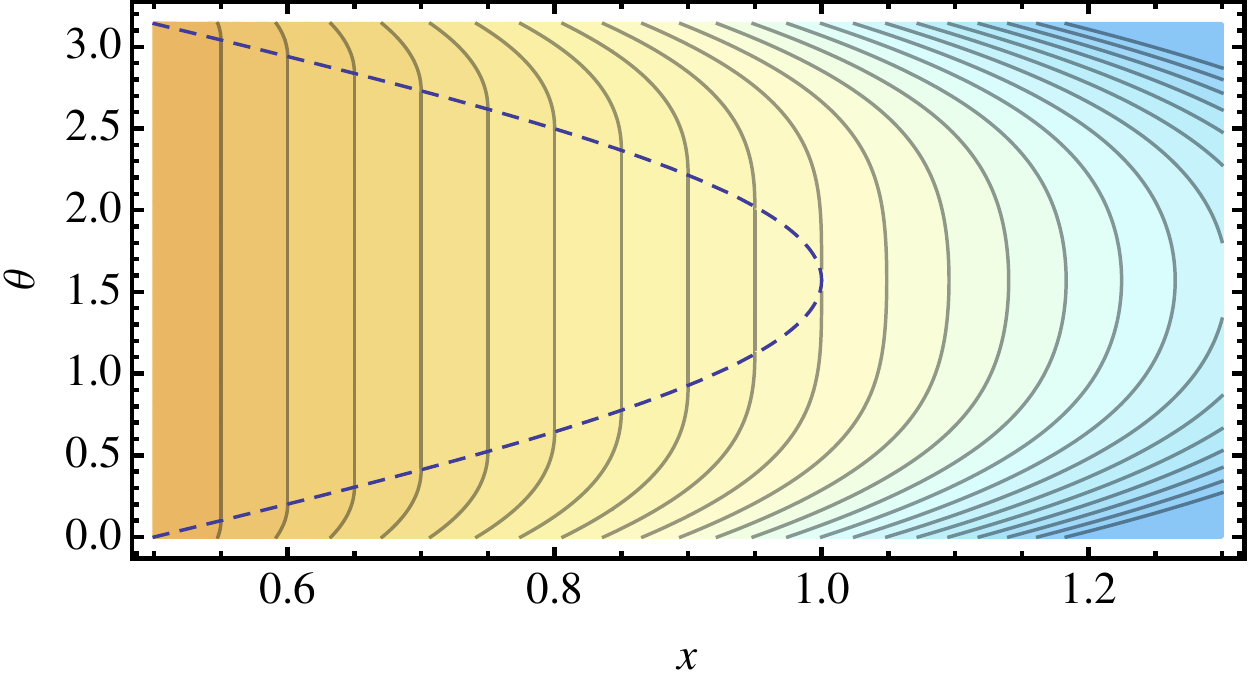} 
\caption{Contour plot of $\beta(x,\theta)$ in a region outside the strip, $x>1/2$. The contour curves are equally spaced in $\beta$. Warmer colors indicate higher temperatures. As before, the dashed lines separate the two regimes $k>2$, and $k<2$. The reflected contour plot is obtained in the region $x<-1/2$.}
\label{tempout}
\end{center}
\end{figure}

\subsubsection*{k<2}
For $k<2$ we do the same calculation but use (\ref{para1}) for the limit of the coefficients in the equations for $S$ and $\bar{S}$. The result is the equation
\be
-4 \,k^2  + 
 k \,(X^2\, (1 + 4 y^2) - 8 \,x\, y\, X\,  Y + 4 \,Y\, (1 + x^2\, Y))=0 \,.
\ee
This equation is again quadratic on $X$ and $Y$, but in contrast to (\ref{dyp}), it depends on $k$.
To complete the system of equations we must also use eq. (\ref{suple}). Combining these equations a quartic equation can be obtained for $X$ or $Y$ alone. We again have four solutions, two of which are always real and the other two are related to each other by complex conjugation. It can be checked that the vectors $\vec{A}$, $\vec{B}$, defined by this system through (\ref{vectorial}) are indeed gradients and the system provides a non trivial solution of the eikonal equations.

The solutions of the quartic equations can be treated numerically. We show an example of the orthogonal coordinate system $a$, $b$ for $k=1/2$ in figure \ref{eiko}. We see the range of the solution extends further than the case $k>2$, but it is again limited to certain bounded region of the plane containing the end points of the intervals. In the limit of $k\rightarrow 0$  we get the massless case shown in figure \ref{eiko2}. In this limit the range extends to the full plane.

On the interval at $y=0$ the solutions are
\bea
B_x &=&\pm
\frac{2 \sqrt{
  2 - k (1 - (4 + k) x^2 + 4 k x^4) + \sqrt{
   4 - k (4 - k) (1 - 4 x^2)}}}{1 - 4 x^2}\,,\\   
  A_y &=& \frac{2 + \sqrt{4 - k (4 - k) (1 - 4 x^2)}}{1 - 4 x^2}\,,
\eea
and are valid in the range $-\sqrt{\frac{2}{k}}<x<\sqrt{\frac{2}{k}}$, outside of which $B_x$ turns out imaginary and the eikonal solution vanish.

Written in term of the angle, we have the following  expression for $\beta=(2\pi)/A_y$,
\be
\beta(x,\theta)=
 \frac{\pi (1 - 4 x^2)}{1 + 
  \sqrt{1 + \frac{
   16 (1 - 
      4 x^2) ( \sin(\theta) - (1 - 
         4 x^2) ) \sin(\theta)(1- \sin(\theta))}{(1 - 8 x^2 + \cos(2 \theta))^2}}}\,.
\ee
This is valid exactly where the solutions for $k>2$ are not, that is, 
\be
0\le \theta\le \acos(2 \sqrt{|x| (1 - |x|)})\,,\hspace{.7cm}  \acos(-2 \sqrt{|x| (1 - |x|)})\le \theta \le \pi\,.
\ee

We see the temperatures are angle dependent for a fix $x$. 
A contour plot of the function $\beta(x,\theta)$ is shown in figure \ref{temp}. The two regimes $k>2$, $k<2$, separated by a dashed curve. Even if there is a "phase transition" in the 
solution, $\beta(x,\theta)$ and its first derivative are continuous at the transition point (while the second derivatives are not). A contour plot for $\beta(x,\theta)$ for the exterior region $\bar{V}$ given by $|x|>1/2$ is shown in figure \ref{tempout} (only the part $x>1/2$ is plotted).

 For angle $\theta=0$ the temperatures are given by the $d=2$, zero mass, solution, $\beta=2\pi (1/4-x^2)$. For angle $\theta=\pi/2$ the solution with $k>2$ applies for all $x$, and the temperatures are like Rindler temperatures corresponding to the plane that is nearer to $x$. For intermediate angles $\beta$ lies within these two curves, see figure \ref{curvas}. This is in accordance with the monotonicity property of relative entropy. This property implies that $\beta(x,\theta)$ for fixed $x,\theta$ has to increase if the region is increased. Hence all $\beta$ of the wall have to be less than the ones of a Rindler half-space including the wall and with boundary coinciding with one of boundaries of the wall. This bound is saturated for angle $\pi/2$. On the other hand, the wall contains a sphere of diameter $L=1$, and $\beta$ for the sphere is the same as the one for a $d=2$ interval. This lower bound is saturated here for angle $\theta=0$. A wall will also contain smaller walls. The corresponding inequality is satisfied because of the property of convexity of the curves shown in figure \ref{curvas}.    
 
\section{Final remarks}

We have calculated from first principles the relative entropy of a localized high energy excitation and the vacuum state in a region $V$ for free fields. The result is given in terms of  a particular geometric problem involving two orthogonal gradient vector fields of the same modulus sourced at the boundary $\partial V$ of the region. The result is the same for scalars and fermions, and it implies the relative entropy is proportional to the excitation energy. The coefficient is an inverse temperature that is generally direction dependent. We computed these temperatures  explicitly for the geometry of a wall.

     \begin{figure}[t]
\begin{center}
\includegraphics[width=0.47\textwidth]{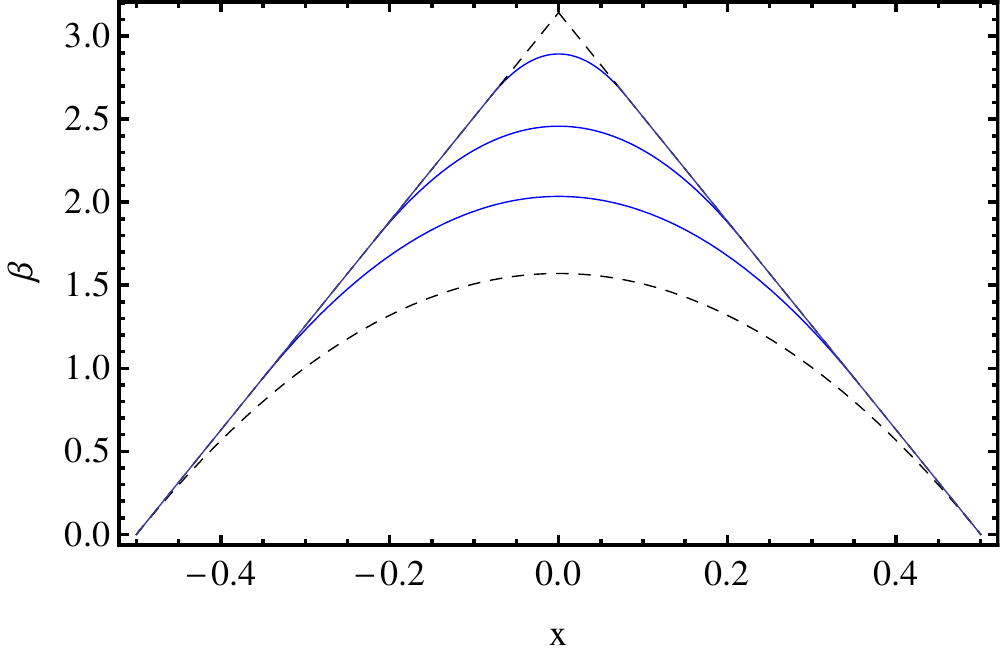} 
\caption{The blue curves are $\beta(x,\theta)$ as a function of $x$ for, from bottom to top,  $\theta=0.3, 0.6, 1$. They are straight lines for  values of $|x|\sim 1/2$, until some specific $|x|$ depending on the angle. For smaller $|x|$ they follow a different analytic shape described in the text. The inverse temperature $\beta$ increases with angle for $\theta\in(0,\pi/2)$. It is always bounded below by the $d=2$ solution $2\pi (1/4-x^2)$ corresponding to $\theta=0$ (lowest black dashed curve)  and bounded above by the $\theta=\pi/2$ solution corresponding to the Rindler solution of the nearest plane (upper black dashed curve). }
\label{curvas}
\end{center}
\end{figure}

One could wonder the reason for the proportionality of relative entropy and energy for localized excitations. This proportionality holds even if the local terms in ${\cal H}$ are not given exclusively by the energy density operator in the general case.
As explained in \cite{nosotros} this is a consequence of the Rindler result plus relative entropy monotonicity. In the Rindler case one can trace this proportionality to the fact that the modular Hamiltonian is a space-time symmetry and the symmetry generator is linear in the energy density operator. In the general case, if we forget about monotonicity and interrogate our explicit calculations, we see the origin of linearity with energy in the fact that even if there is no rotational symmetry in Euclidean space the only source of the local terms in the modular Hamiltonian are the vectors $\vec{A}$, $\vec{B}$ that have a geometric origin. Being gradients of a phase, have specific dimension $1$. These can be thought of as  momentum variables generated by the "rotational" source located at $\partial V$. 

The eikonal equations are a certain extension to higher dimensions of the holomorphicity property that holds in $d=2$. It would be interesting to understand these equations better from the mathematical point of view. In particular, there should be a way to prove the monotonicity of $\beta(\hat{p},\vec{x})$ with the size of $V$ that must hold because of the relation with relative entropy. We would also like to understand how to solve these equations in more general cases. There are interesting properties of these equations such as that if $\alpha$ is a solution, $f(\alpha)$, with $f$ an analytic function, is also a solution (locally). However, this is not enough to produce new interesting solutions to the present problem. The equations are local in nature but the solutions depend strongly on the boundary conditions far away. One might be inclined to think there should be some higher dimensional analogous to the Cauchy integral formula that gives the general solution in $d=2$. That would allow to get the solution from the boundary conditions directly.  

Another open question is how these results modify for interacting theories. Since the local temperatures are by definition a property of the high energy limit, it is reasonable to think the problem will not be modified for theories with free UV fix point, specially super-renormalizable theories. This points towards a universality of the local temperatures. For theories interacting in the UV the situation is much less clear. In this sense, it would be interesting to understand if there is a simple way to compute these temperatures for holographic theories.   

\section*{Acknowledgments}
We thank useful discussions with Gonzalo Torroba.  This work was partially supported by CONICET, CNEA and Universidad Nacional de Cuyo, Argentina. H.C. acknowledges support from an It From Qubit grant of the Simons Foundation.

\appendix

\section{Uniqueness of the solutions given the behaviour at $\partial V$}
\label{ap1}

In this Appendix, we show the solution of the wave equation with multiplicative boundary conditions on the cut $V$ is uniquely determined by its boundary behaviour on $\partial V$. We will treat the case of a scalar, but analogous results hold for the Dirac field.

In order to show this, let us consider a Green function $G(x,y)$ satisfying the boundary conditions of the problem,
\bea
(-\nabla_x^2+m^2) \,G(x,y)&=& \delta^{(d)}(x-y) \, ,\\
\lim_{y_0\rightarrow 0^+} G(x,(y_0,\vec{y}))&=&e^{ 2 \pi s} \lim_{y_0\rightarrow 0^-} G(x,(y_0,\vec{y}))\,, \hspace{0.5 cm} \forall \vec{y}\in V\,, \label{sist}\\
\lim_{|y|\rightarrow \infty} G(x,y)&=&0 \, ,\\
G(x,y) && \textrm{bounded for}\,\, y\rightarrow \partial V\,.  
\eea 
The Green function will not be unique owing to the existence of solutions of the homogeneous wave equation with the same boundary conditions.\footnote{This is in contrast with the case where $s$ is imaginary \cite{scalar}. For imaginary $s$ there are no solutions of the homogeneous wave equation bounded in $\partial V$, and the Green function is unique.} However, in the following we just need any Green function for the problem. Note (\ref{sist}) has a factor $e^{ 2 \pi s}$ opposite to the one in (\ref{bcs}) and (\ref{pin}).

Consider the current 
\be
J_{\mu}^{x}(y)= \partial^y_\mu G(x-y) S(y)-G_s(x-y) \partial_\mu S(y) \, .
\ee
This differ to the current in (\ref{current}) in that now, the Green function satisfies the multiplicative boundary conditions. 
We have 
\be
\partial_\mu J_\mu^{x}(y)= -\delta^{(d)}(x-y) S(y) \,.
\ee
We integrate this equation on $\mathcal{M}$ and notice that due to the opposite factors in the boundary conditions for $S(y)$ and $G(x,y)$ the current is continuous across the cut and there is no boundary value coming from $V$. However, we still have a contribution on $\partial V$, where the current is singular. Taking a thin tube-like surface $\delta V_\epsilon$ of width $\epsilon$ around $\partial V$ we have
  \bea
-S(x) &=& \int_\mathcal{M} d^d y \, \partial_\mu J_\mu^{x}(y)  = -\int_{\partial V_\epsilon} d^{d-1}y \, \eta_\mu(y) J_\mu^{x}(y)\,, \label{rrr}
\eea
where $\eta_\mu(y)$ is the outward pointing unit vector normal to $\partial V_\epsilon$. This shows the full solution $S(x)$ is determined by the values in the limit $x\rightarrow \partial V$. To see this more explicitly, we take local coordinates $y_\parallel,z,\bar{z}$ near a point of $\partial V$, where $y_\parallel$ describes the coordinates along $\partial V$ and the two dimensional complex coordinates $z,\bar{z}$ the directions perpendicular to it. The leading terms in the solutions for $G(x,y)$ and $S(y)$ must have the following general form near a point $y_\parallel\in \partial V$ ($|z|,|\bar{z}| \ll 1$)
\bea
 G(x,y)&\sim & U_1(x,y_\parallel)\, z^{-i s} +U_2(x,y_\parallel)\, \bar{z}^{i s}\,,\\ 
 S(y)&\sim & V_1(y_\parallel)\, z^{i s} +V_2(y_\parallel)\, \bar{z}^{-i s}\,.\label{v}
\eea  
Then we get from (\ref{rrr})
\be
S(x)=-i 4 \pi s \int_{\partial V} dy_\parallel\, \sqrt{g(y_\parallel)} \, (U_1(x,y_\parallel) V_1(y_\parallel)-U_2(x,y_\parallel) V_2(y_\parallel))\,. \label{dds} 
\ee
This shows explicitly how the solution is determined by the asymptotic values on $\partial V$.
However, the functions $V_1$ and $V_2$ are not independent. The solutions are parametrized by only one function of the boundary instead. To see this we can take another solution $\tilde{S}$ of the problem corresponding to the opposite value $-s$, with asymptotic behaviour
\be
\tilde{S}(y)\sim  W_1(y_\parallel)\, z^{-i s} +W_2(y_\parallel)\, \bar{z}^{i s}\,.
\ee
We have, following the above steps but for the current generated by $S$ and $\tilde{S}$,
\be 
0= \int_{\partial V} dy_\parallel\, \sqrt{g(y_\parallel)} \, (W_1(y_\parallel) V_1(y_\parallel)-W_2(y_\parallel) V_2(y_\parallel))\,. \label{owing}
\ee  
Once all these relations are satisfied, (\ref{dds}) gives a unique solution given the boundary asymptotic values for $S$ since the ambiguities on the Green function are due to additions of homogeneous solutions that do not contribute to this formula owing to (\ref{owing}).

In particular, the time reflected, conjugated function $\tilde {S}(x)=S^*(-x_0,\vec{x})$ is a solution of the problem with parameter $-s$. The asymptotic behaviour follows from (\ref{v}) by replacing $z\leftrightarrow \bar{z}$ and taking complex conjugates: $W_1=V_1^*$, $W_2=V_2^*$. Then for any solution $S$ we have the self-consistency relation for the asymptotic values  
\be 
\int_{\partial V} dy_\parallel\, \sqrt{g(y_\parallel)} \, ( |V_1(y_\parallel)|^2-|V_2(y_\parallel)|^2 )=0\,. 
\ee

\section{Massive scalar with boundary conditions on an interval}
\label{ap2}

In this Appendix we find linear partial differential equations of first order for the functions $S(x)$ giving the eigenvectors of the correlator kernels for a massive scalar in an interval in $d=2$. The constant coefficients appearing in these equations are given in terms of solutions of Painlev\'e V equations. The methods we use are an adaptation of the ones in \cite{myers,scalar}. The main difference of these works with the present case is that here the factor in the boundary condition is real (\ref{bcs}), instead of a phase as in \cite{scalar}. When the factor is a phase there are no solutions bounded on the cut, while there are exactly two solutions for a real factor.\footnote{However, with sufficient care, the results of this appendix could be obtained from \cite{scalar} by analytic continuation from imaginary to real $s$. The present function $S$ is then proportional to the one called $S_1$ in that work, where one replaces $a\rightarrow -i s$. Note also that the names of the intervals end-points are interchanged here with respect to \cite{scalar}.}    

We take an interval $\left[L_{1},L_{2}\right]$ with length $L=L_{2}-L_{1}$. We write the solution $S\left(x,y,L_{1},L_{2},m,s\right)$ in complex coordinates as $S\left(z,\bar{z},L_{1},L_{2},m,s\right):=S\left(z\right)$. As shown in Appendix \ref{ap1}, for each $s$ there are exactly two solutions of (\ref{ekg}) and (\ref{bcs}) for an interval. These are complex conjugate of each other, $S(z)$, $\bar{S}(z)$.   
Because $S$ is bounded near $L_1$, and solves the Klein-Gordon equation and boundary conditions, it must have an asymptotic expansion of the form 
\be
S(z)  =  \left(L_{1}-z\right)^{-i s}\,\sum_{k,n=0}^\infty A_{kn}\left(L_{1}-z\right)^k \left(L_{1}-\bar{z}\right)^n 
  +\left(L_{1}-\bar{z}\right)^{i s} \,\sum_{k,n=0}^\infty B_{kn}\left(L_{1}-z\right)^k \left(L_{1}-\bar{z}\right)^n \,, 
  \ee
 where the coefficients are functions of $L$, $s$, and $m$. We will write $A_{k,n}\left(L\right)$ and $B_{k,n}\left(L\right)$. 
We will only need this expansion up to quadratic terms. For convenience, since we have exactly two independent solutions conjugated to each other, we can choose 
\be
A_{00}=1\,,\hspace{.8cm} B_{00}=0\,,
\ee
 which is the solution asymptotically analytic at $L_1$. In Consequence $\bar{S}$ has $\bar{A}_{00}=0$, $\bar{B}_{00}=1$ (we write $A_{00}^*$ for the complex conjugate and $\bar{A}_{00}$ for the quantity belonging to the conjugated solution).  

The Helmholtz (Euclidean Klein-Gordon) equation for $S$ implies relations for the coefficients.
In particular we have $A_{01}=A_{02}=0$, $B_{11}=B_{10}=B_{20}=0$, and
\be
A_{11}=\frac{m^{2}}{4\left(1-is\right)}\;\Rightarrow\;\partial_{L}A_{11}=0\,.
\ee

We have a similar expression for the expansion of $S$ near $L_{2}$
\be
S(z)  =  \left(z-L_{2}\right)^{i s}\,\sum_{k,n=0}^\infty C_{kn}\left(L_{1}-z\right)^k \left(L_{1}-\bar{z}\right)^n
 +\left(\bar{z}-L_2\right)^{-i s} \,\sum_{k,n=0}^\infty D_{kn}\left(L_{1}-z\right)^k \left(L_{1}-\bar{z}\right)^n \,. 
  \ee
and equivalent relations for the coefficients. For the lowest ones we have $C_{01}=C_{02}=D_{10}=D_{20}=0$,
\be
C_{11}=\frac{m^2 C_{00}}{4(1+i s)}\,,\hspace{.7cm} D_{11}=\frac{m^2 D_{00}}{4(1-i s)}\,.
\ee
\subsection{Equations for $S\left(z\right)$ and the coefficients}
A partial derivative $\partial_{L_{1}}S$ also obeys the wave equation and boundary condition, except that it is now unbounded at $L_1$. This divergence can be compensated by adding derivatives of $S$ and $\bar{S}$ with respect to $z$, to obtain a bounded function. This must then be a linear combination of $S$ and $\bar{S}$ by the uniqueness of the solutions. The coefficients on these combinations can be adjusted to be the same at leading order near $L_1$, giving the following equation valid in the plane
\be
D_{00}^{*}\,\partial_{L_{1}}S+D_{00}^{*}\,\partial_{z}S-C_{00}\,\partial_{z}\bar{S}  =  \left(1-is\right)B_{01}^{*}C_{00}\,S+\left(1+is\right)B_{01}D_{00}^{*}\,\bar{S}\,.
\ee
Looking at the expansion of this equation at $L_2$ we get the following differential equations for the coefficients
\bea
\dot{C_{00}}D_{00}^{*} &=& \left(1+is\right)C_{10}D_{00}^{*}-\left(1+is\right)C_{00}D_{01}^{*}-\left(1-is\right)B_{01}^{*}\left(C_{00}\right)^{2}-\left(1+is\right)B_{01}\left(D_{00}^{*}\right)^{2}\,,\\
\dot{D_{00}}D_{00}^{*} &=& -\left(1-is\right)B_{01}^{*}C_{00}D_{00}-\left(1+is\right)B_{01}C_{00}^{*}D_{00}^{*}\,,\\
\dot{D_{01}}D_{00}^* &=&-\frac{1}{4(1-i s)}\left(C_{00} (m^2 C_{00}^* -4 (i+s)^2 B_{01}^* D_{01} +D_{00}^* (4(1+s^2) B_{01}C_{10}^*-m^2 D_{00})  )             \right)\,.
\eea

In the same line, one can notice a rotation operator around the point $L_1$
\be
\partial_{R}^{L1} :=  -i\left(x\partial_{y}-y\partial_{x}-L_{1}\partial_{y}\right)=-\left(L_{1}-z\right)\partial_{z}+\left(L_{1}-\bar{z}\right)\partial_{\bar{z}}=\left(z-L_{2}\right)\partial_{z}-\left(\bar{z}-L_{2}\right)\partial_{\bar{z}}+L\left(\partial_{z}-\partial_{\bar{z}}\right)\,,
\ee
commutes with $-\nabla^2+m^2$, and then $\partial_{R}^{L1}S$ obeys the same equations of motion and boundary conditions as $S$. The divergences and finite leading order terms at $L_1$ can be compensated conveniently, to get the equation 
\be
D_{00}^{*}\,\partial_{R}^{L_{1}}S+L\,D_{00}^{*}\,\partial_{\bar{z}}S-L\,C_{00}\,\partial_{z}\bar{S} =  \left\{ L\left(1-is\right)B_{01}^{*}C_{00}-isD_{00}^{*}\right\} \,S-L\left(1+is\right)B_{01}D_{00}^{*}\,\bar{S}\,,\label{angular}
\ee
which is valid in the plane. Note this is a first order partial differential equation for $S$ and $\bar{S}$.
Evaluating the coefficients of the expansion of this equation at $L_1$ and $L_2$ up to first order give the following algebraic equations for the coefficients
\bea
0 &=& L\left(-1+is\right)B_{01}^{*}C_{00}D_{00}+L\left(1+is\right)B_{01}C_{00}^{*}D_{00}^{*}+2is\,D_{00}D_{00}^{*}\,,\\
0 &=& -\left(i+s\right)\left(i+2s\right) D_{00}^{*}D_{01}-L\frac{m^{2}}{4}\left(\left|D_{00}\right|^{2}-\left|C_{00}\right|^{2}\right) + L\left(1-is\right)^{2}B_{01}^{*}C_{00}D_{01}\nn\\
&&\hspace{8.5cm}-L\left(1+s^{2}\right)B_{01} C_{10}^{*}D_{00}^{*}\,.\\
0 &=& L (m^2 - 4 (1 + s^2) B_{01} B_{01}^*) C_{00} - 
 4 (-i + s) ((i + L (-i + s) A_{10}^*) B_{01} - 
    L (-2 i + s) B_{02}) D_{00}^*\,,\\
0 &=&  -4 (i + s) A_{10} (L (i + s) B_{01}^* C_{00} - i D_{00}^*) + \nn
 L (4 (-2 + 3 i s + s^2) B_{02}^* C_{00}\\ && \hspace{7cm}+ (m^2 - 4 (1 + s^2) B_{01} B_{01}*) D_{00}^*)\,.   
\eea
Using again the same argument, we note that the function $S(z)$, reflected around the mid-point of the interval,
\be
S_R(z)=S(L_1+L_2-\bar{z})\,,
\ee
 is again a solution, and hence a linear combination of $S$ and $\bar{S}$. Evaluating at leading order at $L_1$ we get an equation valid in the entire plane
\be
S_R(z)-D_{00}(L) S(z)-C_{00}(L) \bar{S}(z)=0\,, 
\ee
Expanding this equation at the end points up to second order we get
\bea
A_{10}D_{00}+B_{10}^{*}C_{00} & = & D_{01}\,,\hspace{.7cm} C_{10} D_{00}+C_{00}D_{00}^*=B_{01}\,,\\
B_{01}D_{00}+A_{10}^{*}C_{00} & = & C_{10}\,, \hspace{.7cm} D_{00}=D_{00}^*\,, \\
C_{00}C_{10}^*+D_{00}D_{01} &=& A_{10}\,,\hspace{.7cm}  |C_{00}|^2 = |D_{00}|^2+1\,.
\eea

$D_{00}$ and $C_{00}$ are dimensionless and can be written as functions of the dimensionless parameter $t=m L$. From the above equations we have 
\be 
D_{00}(L)=i \, d(t)\,,\hspace{.7cm}|C_{00}|^2=d^2(t)+1\,.
\ee
$D_{00}$ is purely imaginary, and $d(t)$ real. Combining all equations we get that the function $d(t)$ satisfies the non-linear ordinary differential equation (\ref{painleve0}) of the Painlev\'e V type  quoted in section \ref{slab}.

As explained in \cite{review} the boundary condition for small $t$ (which corresponds to small mass) can be obtained by using (\ref{ints2}) with the massless solution and the Green function in the limit of small mass. The boundary condition for large $t$ follows from the Painlev\'e connection formulae \cite{connection}. In this way we get the equations (\ref{boundary1}) and (\ref{boundary2}) in section \ref{slab}. 

The algebraic and differential equations determine all variables in terms of $d(t)$, $d'(t)$, and $C_{00}(t)$. In particular, the equation (\ref{angular}) takes the form quoted in (\ref{efi1}).

\end{document}